\documentclass[a4paper,12pt]{article} 	
\usepackage[utf8]{inputenc}
\usepackage[english]{babel}
\usepackage{geometry}              	       %serve per i margini altrimenti manualmente \noindent
\usepackage{caption}
\usepackage{amsmath}
\usepackage{float}
\usepackage{amsfonts}
\usepackage{mathtools}
\usepackage{slashed}
\usepackage{placeins}
\usepackage{subfig}
\usepackage{adjustbox}
\usepackage{lineno}
\usepackage{authblk}
\usepackage{color}
\usepackage[absolute]{textpos}
\pdfoutput=1

\DeclareMathOperator{\Tr}{Tr}

\renewcommand*{\Affilfont}{\normalsize\normalfont}
    
\newsavebox\affbox
	\date{}

\title{\Huge New extended interpolating fields built from three-dimensional fermions \vspace{1cm}}
\author[1]{Mauro Papinutto}
\author[1]{Francesco Scardino}
\author[2]{Stefan Schaefer}

\affil[1]{\Affilfont 	Dipartimento di Fisica, "Sapienza" Università di Roma, and INFN, Sezione di Roma, Piazzale Aldo Moro 2, I-00185 Roma, ITALY}

\affil[2]{	Neumann Institute for Computing, DESY, Platanenallee 6, 15738 Zeuthen, Germany}
\date{}    

\begin{document}
	
	% make the title area
	\maketitle
\begin{textblock}{5}(10,0.5)
	\null\mbox{}\hfill{\tt DESY 18-091 }
\end{textblock}

	\begin{abstract}
		New extended interpolating operators made of quenched three
		dimensional fermions are introduced in the context of lattice QCD. The mass of
		the 3D fermions can be tuned in a controlled way to find a better overlap of
		the extended operators with the states of interest. The extended operators have
		good renormalization properties and are easy to control when taking the
		continuum limit. Moreover the short distance behaviour of the two point
		functions built from these operators is greatly improved with respect to Jacobi smeared sources and point sources.
		A numerical comparison with point sources and Jacobi smeared sources on 
		dynamical $2+1$ flavour configurations is presented.
	\end{abstract}

\section{Introduction}
	A serious problem which limits the attainable precision in lattice QCD
	computations of hadronic observables is the exponential suppression of the
  signal-to-noise ratio in Euclidean time \cite{Parisi:1983ae}. It is 
  therefore important to reduce the coefficient of this deterioration 
  as much as possible.\\
	One first step is to find improved interpolating
	operators, which allow to decrease statistical and systematic 
  errors in the extraction of the hadron spectrum, form factors 
  and matrix elements. The guiding principle here is twofold:
  on the one hand, one wants to enhance the coupling to the ground
  state with respect to the excited states. On the other hand, 
  such a method should not increase the statistical noise encountered.\\
  While point sources do in principle suffice to excite the target hadrons in
  the two-point functions, it is an old idea that the overlap with the
  ground state can be improved by using extended sources. This is essential in
  the face of a signal-to-noise ratio which deteriorates with growing Euclidean time
  distances, allowing for plateau regions (from which the masses
  and other physics quantities are extracted) to start at earlier times. In
  particular smoothing techniques based on the iterated application of the
  three-dimensional Laplace
  operator \cite{Gusken:1989ad,Alexandrou:1990dq,Allton:1993wc} have shown to be
  successful in many applications. They provide a way to change
  the relative couplings of the various states which contribute to a given
  two-point function. This can be exploited by setting up a generalized
  eigenvalue problem (GEVP) to get a better handle of the
  ground state\cite{Luscher:1990ck}. \\
  A drawback of these smoothing techniques is that they are quite
  empirical and it is difficult to predict the relevant parameters as
  the lattice spacing is changed. While being computationally rather
  economical in general, their iterative nature can also
  lead to significant cost if a certain smoothing radius has to be created
  on a fine lattice.\\
	Therefore, we here propose an alternative construction of such extended 
  interpolating operators which is based on (quenched) 3D fermions.  The big advantage 
  of this construction is that it can be formulated  in
  terms of a field theory, such that some level of theoretical control can be
  gained. We will consequently be able to show that these fermions are well 
  behaved under renormalization and in fact improve the short distance behaviour
  of two-point functions. Since their construction involves the solution
  of a linear system of equations, the vast literature and improvements
  in this area can make them also computationally economical.\\
  Since such a construction is for quark fields, but it is the resulting
  hadrons which we measure on the lattice, it is important to state how
  the composite hadrons are constructed. To this end, we review for completeness  
  the classification of the octet and decouplet baryonic
	interpolating operators which we have classified according to the irreducible
	representations of the cubic group and we have confronted with the ones obtained 
  in \cite{Basak:2005aq}\cite{Basak:2005ir}. This
	classification has the advantage of being exhaustive by construction.\\
  The outline of the paper is therefore as follows: we first give the baryonic (point) operators
  which we will consider in Sec.~\ref{s2}. The three-dimensional fermions are defined in Sec.~\ref{s3}
  and these two ingredients are put together in Sec.~\ref{s4}, where the extended baryon sources
  are defined. After that, we 
	 compare them in a typical numerical simulation with point sources and Jacobi smeared sources, also
   with respect to the question of the performance in a GEVP.

  \section{Classification of baryonic operators\label{s2}}
	Baryons are bound states of three valence quarks. The interpolating operators $\mathcal{B}$ need to have a well defined set of quantum numbers such that the corresponding Hilbert space operator $\hat{\mathcal{B}}$ projects onto the state we are interested in.
	On the lattice, rotational symmetry breaks down and is replaced by the cubic group $SO(3,\mathbb{Z})$  whose elements are the matrices of $SO(3)$ with integer entries. In order to correctly classify the operators, we will use the spin covering group  $Spin(3,\mathbb{Z})$ of  $SO(3,\mathbb{Z})$. In fact the $Spin(3)$ group is isomorphic to $SU(2)$ and  allows us to work with the continuum Weyl spinor notation on the lattice.\\
	A generic gauge invariant three-quark operator has the form
	\begin{equation}
	\mathcal{B}(x) = u^{a}_{\alpha}(x)d^{b}_{\beta}(x)s^{c}_{\gamma}(x)\ t^{\alpha\beta\gamma}\epsilon_{abc},
	\end{equation}
	where $u,d,s$ are the quark fields, which are in the fundamental representation of $SU(3)_c$. Furthermore, despite what is suggested by the notation, the fields have a definite but not necessarily different flavour. The tensor $t^{\alpha \beta \gamma}$ depends on the $SO(3,\mathbb{Z})$ representation the baryon falls in,  $\epsilon_{abc}$ is the only allowed invariant colour tensor. Greek indices $\alpha,\beta,\ldots$ are Dirac indices while the Latin ones $(a,b,\ldots)$ represent colour.\\
	In order to classify the baryonic operators according to the irreducible representations of the $SU(3)$ flavour group and the rotation group on the lattice, we need  to classify the tensors $t^{\alpha\beta\gamma}$ introduced above according to the irreducible representations of the spin covering $Spin(3,\mathbb{Z})$ of the cubic group $SO(3,\mathbb{Z})$. For spin $\frac{1}{2}$ and spin $\frac{3}{2}$ baryons, it can be proven \cite{Johnson:1982yq} that there is a one-to-one correspondence between the corresponding $Spin(3,\mathbb{Z})$ representations on the lattice and the continuum ones. We thus use the continuum notation of dotted and undotted Weyl spinors in the following.\\
	This brings us to the operators that have been actually used in the simulations. Consider the case in which two flavours are equal, namely for spin $s=\frac{1}{2}$ the case of the nucleon. It is possible to show that there are two nucleon operators $\hat{N}$ and $\tilde{N}$ in the $\left(\frac{1}{2},0\right)$ representation 
	\begin{align}
	\label{proton12}
	\nonumber
	&\hat{N}^{(\frac{1}{2},0)}_{\frac{1}{2},\frac{1}{2}}=\frac{1}{\sqrt{2}}(u_1 d_2-u_2 d_1)u_1\qquad\qquad \hat{N}^{(\frac{1}{2},0)}_{\frac{1}{2},-\frac{1}{2}}=\frac{1}{\sqrt{2}}(u_1 d_2-u_2d_1)u_2\\
	&\tilde{N}^{(\frac{1}{2},0)}_{\frac{1}{2},\frac{1}{2}}=\frac{1}{\sqrt{2}}(u_{\dot{1}}d_{\dot{2}}-u_{\dot{2}}d_{\dot{1}})u_1\qquad\qquad \tilde{N}^{(\frac{1}{2},0)}_{\frac{1}{2},-\frac{1}{2}}=\frac{1}{\sqrt{2}}(u_{\dot{1}}d_{\dot{2}}-u_{\dot{2}}d_{\dot{1}})u_2
	\end{align}
	and one in the $\left(\frac{1}{2},1\right)$ representation 
	\begin{align}
	\label{proton3}
	\nonumber
	&{N}^{(\frac{1}{2},1)}_{\frac{1}{2},\frac{1}{2}}=\frac{1}{\sqrt{14}}\{(u_{\dot{1}}d_{\dot{2}}+u_{\dot{2}}d_{\dot{1}})u_1-2u_{\dot{1}}d_{\dot{1}}u_2+2(u_{\dot{1}}d_2-u_{\dot{2}}d_1)u_{\dot{1}}\}\\ &{N}^{(\frac{1}{2},1)}_{\frac{1}{2},-\frac{1}{2}}=-\frac{1}{\sqrt{14}}\{(u_{\dot{1}}d_{\dot{2}}+u_{\dot{2}}d_{\dot{1}})u_2-2u_{\dot{2}}d_{\dot{2}}u_1+2(u_{\dot{2}}d_1-u_{\dot{1}}d_2)u_{\dot{2}}\}.
	\end{align}
	In the case of three equal flavours, i.e. the $\Omega$ baryon, there are two $s = \frac{3}{2}$ decouplet operators
	\begin{align}
	\label{omega12}
	\nonumber
	&{\Omega}^{(\frac{1}{2},1)}_{\frac{3}{2},\frac{3}{2}}=s_{\dot{1}}s_{\dot{1}}s_1 \qquad\qquad {\Omega}^{(\frac{1}{2},1)}_{\frac{3}{2},\frac{1}{2}}=\frac{1}{\sqrt{3}}\{s_{\dot{1}}s_{\dot{1}}s_2+s_{\dot{1}}s_{\dot{2}}s_1+s_{\dot{2}}s_{\dot{1}}s_1 \}\\
	\nonumber
	&{\Omega}^{(\frac{3}{2},0)}_{\frac{3}{2},\frac{3}{2}}=s_1s_1s_1 \qquad\qquad {\Omega}^{(\frac{3}{2},0)}_{\frac{3}{2},\frac{1}{2}}=\frac{1}{\sqrt{3}}\{s_1s_1s_2+s_1s_2s_1+s_2s_1s_1 \}\,,\\
	\end{align}
	where we have reported only the spin up components. Here we have shown the $\left(a,b\right)$ representations, the other half $\left(b,a\right)$ are obtained by switching the dotted and un-dotted indices of the $\left(a,b\right)$ representations. In order to have operators with definite transformation properties under the action of parity, we need to consider
	\begin{align}
	\label{parityops}
	&\mathcal{B}^{\pm}_{(a,b)\oplus(b,a)} =\mathcal{B}_{(a,b)}\mp\mathcal{B}_{(b,a)} \ ,
	\end{align}
 	where $\pm$ indicates the parity eigenvalue and $(a,b)$ the irreducible representation. These operators have been employed in the study of the 3D extended interpolating operators, which we are going to define in the sections below. This classification coincides with the work done in \cite{Basak:2005aq}\cite{Basak:2005ir}.

  \section{Definition of the 3D fields\label{s3}}
    We build new extended operators by defining three dimensional fermions living on a single time-slice, coupling them to the physical quarks propagating in $4$ dimensions.\\
	The three dimensional fermion fields $\varphi$ are defined as ordinary spin $1/2$ quark fields constrained to stay on a fixed time-slice. A possible Lagrangian density for these boundary fields is
	\begin{equation}
	\mathcal{L}_{3D}	=\delta(x_0-\tau) \sum_{i=1}^{N_f}\bar{\varphi_i}\left(\slashed{D}+m^i_{3D}\right)\varphi_i
	\end{equation} 
	with $\tau$ the three dimensional space-like hyper-surface at a fixed arbitrary time. 
	It is obvious that the action is
	\begin{equation}
	S_{3D} = \int d^3\mathbf{x}\ \sum_{i=1}^{N_f}\bar{\varphi_i}\left(\slashed{D}+m^i_{3D}\right)\varphi_i\ .
	\end{equation}
	One immediately sees that these fermion fields must have a canonical dimension of one $\left[\varphi\right] = 1$. In order for these objects to remain on the intended surface, we impose the following constraint
	\begin{equation}
	D_0 \varphi_i\big\rvert_{x_0=\tau} = 0
	\end{equation}
	such that these fields do not propagate in time.\\
	This leaves in the $S_{3D}$ action only the spatial part of the Wilson-Dirac operator, which on the lattice we will represent as follows
	\begin{equation}
	\label{3daction}
	S_{3D} = a^3 \sum_{\mathbf{x}} \bar{\varphi}(\mathbf{x})\mathcal{D}\varphi(\mathbf{x}), \qquad \mathcal{D} = \frac{1}{2}\sum_{i=1}^{3}\{\gamma_i(\nabla^{*}_i+\nabla_i)-a\nabla^{*}_i\nabla_i\}+{m_{3D}}
	\end{equation}
	the sum over the flavours is understood as all other non manifest indices. The 3D fermions are quenched and therefore they only provide valence contributions. However they are coupled to the spatial part of the gauge field and therefore receive quantum corrections and need to be renormalized.
	
\section{Extended operators\label{s4}}
	We now proceed to define the three dimensional extended operators. The idea is to couple the $3D$ fields defined above to the ordinary four dimensional quark fields $\psi$ in a way that defines a spatially extended fermion propagating through time. \\
	An extended quark field $q(\mathbf{x},t)$ for a fixed flavour can be defined as follows
	\begin{equation}
	q(\mathbf{x},t) = a^3 \sum_\mathbf{y} \varphi(\mathbf{x})\ \bar{\varphi}\gamma_5\psi(\mathbf{y},t)\ ,
	\end{equation}
	where the fields $\varphi$ are fixed at the time-slice $t$. We see that the 3D fields are coupled via bilinear operators.\\
	 These bilinears are arbitrary and the presence of the $\gamma_5$ comes from the empirical observation that the pseudo-scalar operator provides a better signal. Now, the extended quark propagator, after integrating out both the 3D and 4D fermions, takes the form 
	\begin{equation}
	\label{extendedprop}
	\langle 	q(\mathbf{x}',t')\bar{q}(\mathbf{x},t) \rangle = a^6 \sum_{\mathbf{y},\mathbf{y}'}S_{3D}(\mathbf{x}',\mathbf{y}')\gamma_5S_{4D}(\mathbf{y}',t',\mathbf{y},t)\gamma_5 S_{3D}(\mathbf{y},\mathbf{x})\ .
	\end{equation}
	We see that a $3D$ quark is propagating in space with $S_{3D}(\mathbf{x},\mathbf{y})$, then,it gets transported to the target time slice by the four dimensional propagator where the loop is closed by another three dimensional propagator.\\
	This construction can be immediately extended to meson and baryon operators. Suppose for instance that we want to study a baryonic operator $B$ as given in Eqs. \ref{proton12}-\ref{omega12}. The corresponding 3D extended operator is built as follows: we first define a 3D baryon operator $\boldsymbol{B}$  built from the $\varphi$ fields on a given time slice, with the same quantum numbers of the original operator. Then  we couple it with the appropriate number of 3D-4D bilinears in order to make it propagate through time
	\begin{equation}
	\label{3Dextbaryon}
	O_{B}(\mathbf{x},t) = a^9 \sum_{\mathbf{x_1},\mathbf{x_2},\mathbf{x_3}} \boldsymbol{B}(\mathbf{x})\ \bar{\varphi}\gamma_5\psi(\mathbf{x_1},t)\ \bar{\varphi}\gamma_5\psi(\mathbf{x_2},t)\ \bar{\varphi}\gamma_5\psi(\mathbf{x_3},t)
	\end{equation}
	namely we have a bilinear for every quark that composes the operator $B$. We immediately notice that while a normal baryonic operator has canonical dimension $\left[B\right] = \frac{9}{2}$ the extended operator has a much lower one $\left[O_{B}\right] = \frac{3}{2}$.\\
	The corresponding baryonic two point function is the same as the usual two point function but where the normal four dimensional quark propagators are replaced by the expression in equation (\ref{extendedprop}).\\
	One may wonder if a similar construction could be done employing 3D bosonic fields instead of the fermionic ones. After all building extended operators with a 3D scalar propagator would more closely resemble the Jacobi smeared sources. Unfortunately, as we show in Appendix A, such an approach would not lead to the
	same benefits as the one based on fermions presented here.
	\subsection{Renormalizability}
	We now briefly review the renormalization properties of the $S_{3D}$ action and of 3D extended operators in general.\\
	 To do so we start by listing the symmetries of  $S_{3D}$. The action is invariant under the transformations of the three dimensional Euclidean lattice rotation group $SO(3,\mathbb{Z})$.
	 The vector part of the flavour group $SU_V(N_f)\otimes U_V(1)$ is still a symmetry, while the axial part is broken by the Wilson term.
	 	There is also the "gamma" symmetry
	 \begin{equation}
\varphi(\mathbf{x})\rightarrow e^{i \alpha \Gamma}\varphi(\mathbf{x}), \qquad \bar{\varphi}(\mathbf{x})\rightarrow \bar{\varphi}(\mathbf{x})e^{-i\alpha \Gamma},\qquad U_i(x)\rightarrow U_i(x)
	 \end{equation}
	 where $\Gamma = i \gamma_0 \gamma_5$ is the matrix that commutes with the 3D Wilson-Dirac operator, while the four dimensional fermions are left invariant. Then there is parity
	 \begin{equation}
	\varphi(\mathbf{x})\rightarrow \gamma_0\varphi(-\mathbf{x}), \qquad \bar{\varphi}(\mathbf{x})\rightarrow \bar{\varphi}(-\mathbf{x})\gamma_0,\qquad U_i(x_0,\mathbf{x})\rightarrow U^{\dagger}_i(x_0,-\mathbf{x}-a \hat{i})
	\end{equation} 
	and charge conjugation
		 \begin{equation}
	\varphi(\mathbf{x})\rightarrow -\bar{\varphi}(\mathbf{x})C, \qquad \bar{\varphi}(\mathbf{x})\rightarrow C^{-1}\varphi(\mathbf{x}),\qquad U_i(x_0,\mathbf{x})\rightarrow U^{*}_i(x_0,\mathbf{x})
	\end{equation} 
	where $C$ is the charge conjugation matrix.\\
	Under renormalization the 3D action $S_{3D}$ could mix with the following two and three dimensional operators
	\begin{align}
	\nonumber
		\bar{\varphi}\varphi,\quad \bar{\varphi}\gamma_0\varphi,\quad \bar{\varphi}\gamma_5\varphi,\quad \bar{\varphi}\gamma_0\gamma_5\varphi,\quad \bar{\varphi}\mathcal{D}\varphi,\quad\bar{\varphi}\gamma_0\mathcal{D}\varphi,\quad\bar{\varphi}\gamma_5\mathcal{D}\varphi,\quad\bar{\varphi}\gamma_0\gamma_5\mathcal{D}\varphi\\
	\end{align}
	however using the discrete symmetries we have listed above and the $SO(3,\mathbb{Z})$ symmetry, we see that the only two allowed operators are $\bar{\varphi}\varphi$ and $\bar{\varphi}\mathcal{D}\varphi$ which give the usual mass and field renormalization.\\
	In Appendix B we show that the field $Z_{\varphi}$ and 3D mass $\delta m_{3D}$ renormalizations are $\log$ divergent at one loop like in the 4D case.
	\subsubsection{3D extended operators and short distance structure}
	We now look at the renormalizability of the extended operators themselves and what this implies for their short distance properties.\\
	As we have seen in equation (\ref{3Dextbaryon}), there are two ingredients that make a 3D extended operator $O_{3D}$: the three dimensional baryon/meson interpolating operator and the bilinear operators that couple it to the four dimensional fermions.\\
	The three dimensional interpolating operator that lives on a fixed time-slice has dimension $3$ if it is a baryon and dimension $2$ if it is a meson; mixing at most with operators of the same dimension. It is clear that there are no lower dimensional operators with the same quantum numbers, in both cases, that could mix with our operator of interest (the only exception concerns the singlet scalar meson which can mix with a lower dimensional operator, the identity, associated with a  power divergence $a^{-2}\mathbb{I}$).\\
	This leaves us to understand what happens to the bilinears on a fixed time-slice. We know that a bilinear $\bar{\varphi}\gamma_5\psi$ has no clear transformation properties under parity and charge conjugation, therefore we just add its Hermitian conjugate which does not affect the two point function. Now
	\begin{equation}
	\bar{\varphi}\gamma_5\psi+\bar{\psi}\gamma_5\varphi
	\end{equation}
	has definite transformation properties under parity and charge conjugation, namely ${P=-1}$, ${C=+1}$.  We immediately see that all other possible bilinear operators with some combination of $\gamma$ matrices between the fields will not transform in the same way under these symmetries and rotations. Therefore the bilinear is purely multiplicatively renormalizable.\\
	We now know that the single pieces are well behaved and do not introduce further divergences to the extended operators that were not already in the original interpolating operator and fields. We have to make sure that when everything is combined and summed together nothing bad happens.
	On a fixed time-slice, $t = 0$ for convenience, the worst divergence of an extended operator happens when all the bilinears get close to the $3D$ interpolating operator. Suppose that $\boldsymbol{B}$ is in the origin. Then the operator product expansion (OPE) tells us that
	\begin{equation}
 \boldsymbol{B}(0)\ \bar{\varphi}\gamma_5\psi(\mathbf{x_1})\ \bar{\varphi}\gamma_5\psi(\mathbf{x_2})\ \bar{\varphi}\gamma_5\psi(\mathbf{x_3}) \rightarrow \frac{1}{(\mathbf{x}^2)^{3}}B(0) \quad \mathbf{x_1},\mathbf{x_2},\mathbf{x_3} \rightarrow 0
	\end{equation}
	where $B(0)$ is the four dimensional baryonic operator. Since now there are three $3D$ integrations, the degree of divergence is $9-6=3$ and therefore the singularity is integrable and does not give rise to additional UV divergences.\\
	We conclude that the $3D$ extended operators are well behaved under renormalization and are finite once the fields and  interpolating operators are renormalized.
	\subsection{OPE and short distance behaviour}
\begin{figure}[H]
	\centering
	\subfloat[3D3D-JJ]{\includegraphics[width=0.49\textwidth,height=\textheight,keepaspectratio]{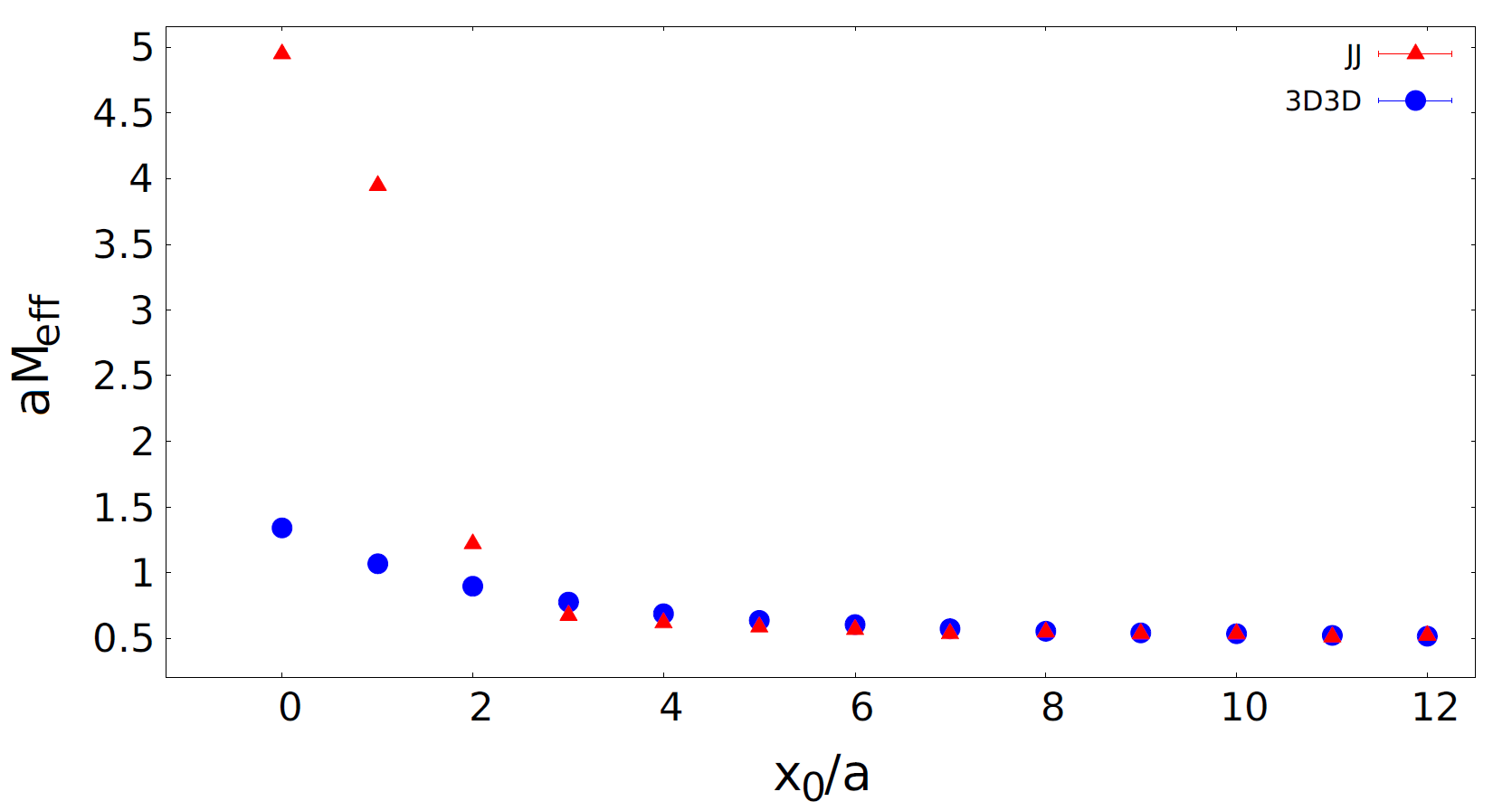}}
	\subfloat[3D3D-JP]{\includegraphics[width=0.49\textwidth]{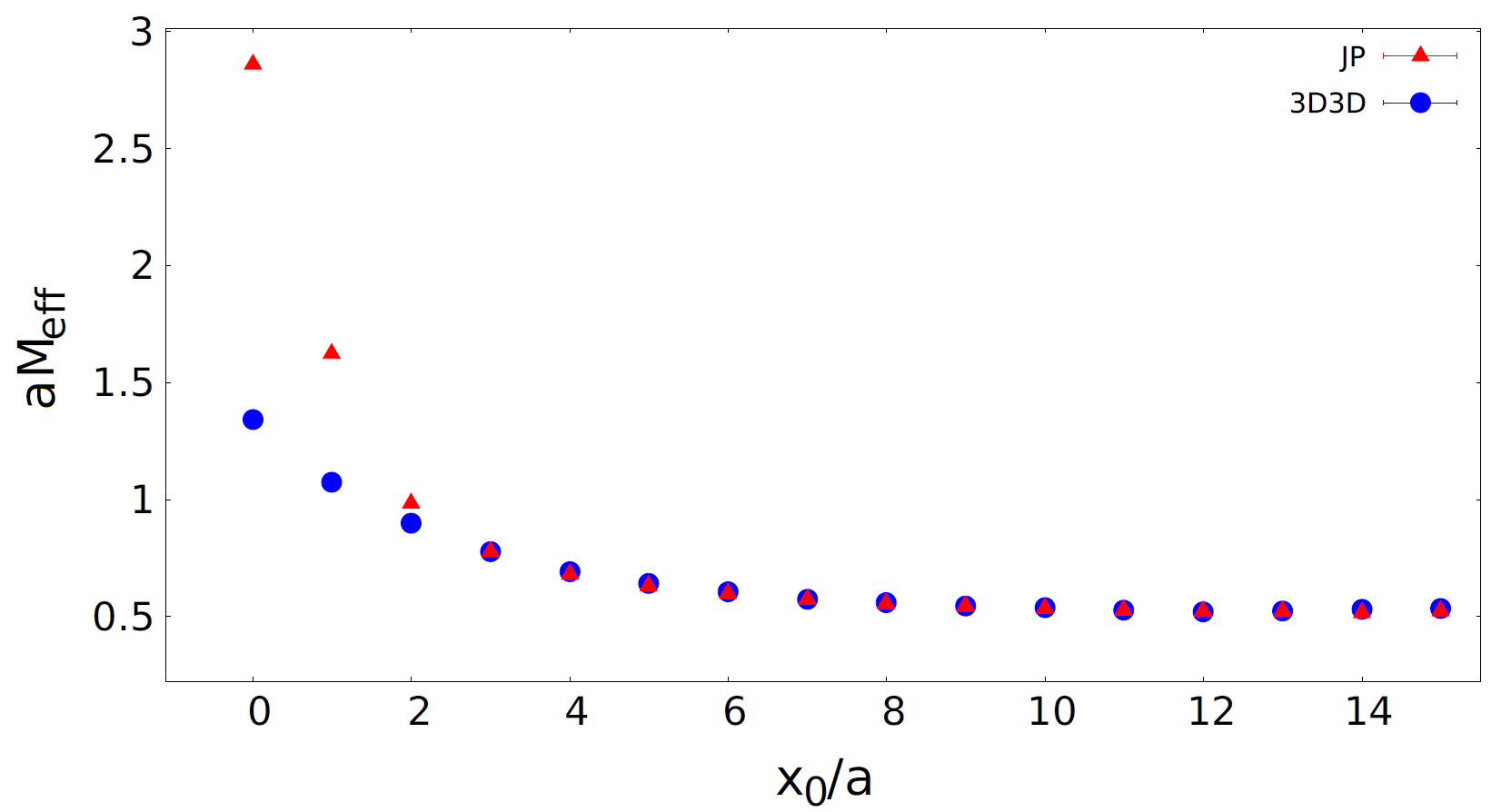}}
	\hfill
	\subfloat[3D3D-PP]{\includegraphics[width=0.49\textwidth,height=\textheight,keepaspectratio]{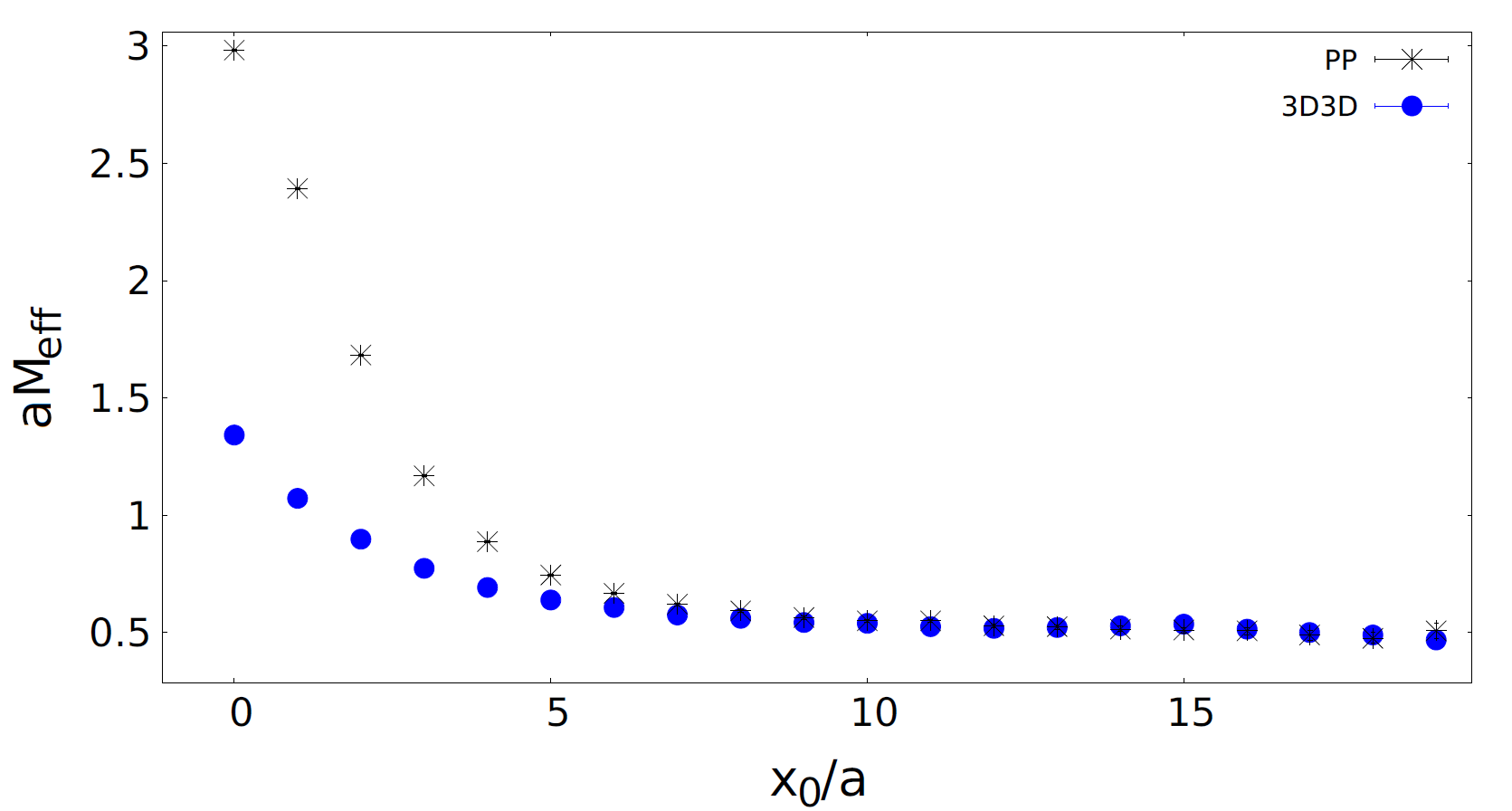}}
	\subfloat[3D3P-JP]{\includegraphics[width=0.49\textwidth]{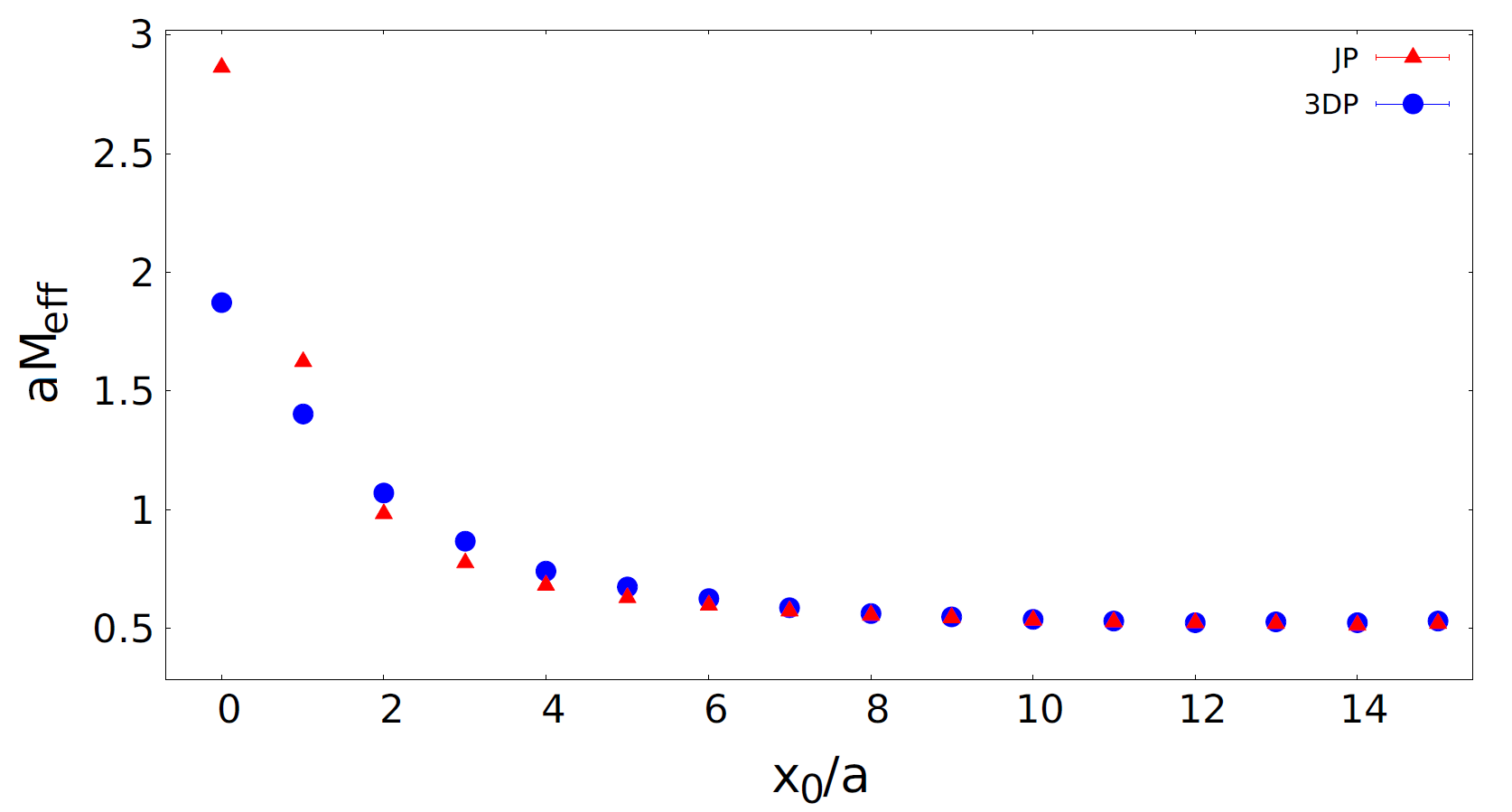}}
	\caption{The comparison between the effective masses of the nucleon calculated for different operators with an emphasis at the short distance behaviours. The notation stands for which operator is in the source and which is in the sink $O_i O_j$ means that $O_i$ is in the source and $O_j$ in the sink. In the plots $3D$ stands for $3D$ extended operator, $J$ for Jacobi smeared source and $P$ for point source. It is possible to see a strong improvement in the short distance behaviour for the 3D extended operators in comparison with Jacobi and point sources.}
	\label{figure_shortd}
\end{figure}

	We have already emphasized that the extended operators have a lower canonical dimension than the original interpolating operator, this obviously affects the short distance behaviour of the correlation functions which can be built from them.\\
	To analyse this behaviour, we will consider a box  of size $T\times L^3$ with spatial periodic boundary conditions in the continuum limit. We know that a baryonic two point function will have dimension $9$, since $\left[B\right] =\frac{9}{2}$ and therefore its behaviour, at zero momentum, for small times is
	\begin{equation}
	C_B(t) \sim \int d^3\mathbf{x}\ \frac{1}{(\mathbf{x}^2+t^2)^{9/2}} \sim \frac{64 \pi}{105 t^6}
	\end{equation}
	so it is badly divergent for small times. On the other hand, a baryonic 3D extended operator has dimension $\left[O_{B}\right] = \frac{3}{2}$ therefore the corresponding two point function behaves much better at short times
	\begin{equation}
	\label{2pextbaryon}
	C_{O_{B}}(t) \sim \int d^3\mathbf{x}\ \frac{1}{(\mathbf{x}^2+t^2)^{3/2}}\sim -8\pi \log \frac{t}{L}\ .
	\end{equation}
	The short distance behaviour is improved from a polynomial to a logarithmic divergence. The improvement of the short distance behaviour of the two point function has been observed from the simulations as it can be seen from Fig. \ref{figure_shortd}, where different methods are compared for the parameters given at the beginning of Sec 5.
	For the mesons the situation is even more dramatic. In fact a normal meson operator has dimension $3$ and therefore its two point function at short distance behaves very badly as
	\begin{equation}
		C_M(t) \sim \int d^3\mathbf{x}\ \frac{1}{(\mathbf{x}^2+t^2)^3}\sim	 \frac{\pi^2}{2t^3}
	\end{equation}
	 on the other hand a 3D extended meson operator has dimension $1$ and its corresponding two point function has no short distance divergences at all
	\begin{equation}
	C_{O_{M}}(t) \sim \int d^3\mathbf{x}\ \frac{1}{\mathbf{x}^2+t^2}\sim 4\pi L .
	\end{equation}
	Another consequence of the reduced canonical dimension of the 3D extended operators is that their OPE involves a lot less operators than there are in the OPE of the usual interpolating operators. In particular the OPE cannot involve gluonic operators, in fact the operator with lowest canonical dimension is $\Tr F_{\mu\nu}F^{\mu\nu}$ which has dimension $4$, while we have seen that the product of two 3D extended baryonic operators has dimension $3$ and $2$ for the mesons.\\
	This fact implies the absence of gluon operators in the OPE of the two point correlation functions of baryonic and meson operators.
	For instance for a 3D extended baryonic operator the OPE is like
	\begin{equation}
	\langle O_{B}(x)\bar{O}_{B}(0) \rangle\sim \frac{1}{x^3}+\frac{m}{x^2}+\frac{1}{x^2} \langle\bar{q}q\rangle +\ldots
	\end{equation}
	where $\bar{q}q$ is the extended quark condensate. And for a 3D extended meson
	\begin{equation}
	\langle O_{M}(x)\bar{O}_{M}(0) \rangle\sim \frac{1}{x^2}+\frac{m}{x}+\frac{1}{x} \langle\bar{q}q\rangle
	\end{equation}
	there can be no further polynomial terms.
	
\section{Numerical results}
The lattice simulations were performed on the  CLS $N_f = 2+1$ gauge configurations \cite{Bruno:2014jqa} that adopt	the open boundary conditions in time \cite{Luscher:2011kk}. The new code is based on {\tt openQCD} \cite{Luscher:2012av}. In the present exploratory study we have used the H101 flavour $SU(3)$ symmetric $96\times32^3$ 
ensemble with	$M_\pi=M_K=420$\,MeV and a  lattice spacing $a = 0.086$\,fm \cite{Bruno:2016plf}.
\begin{figure}[h]
	\centering
	\includegraphics[width=0.5\linewidth]{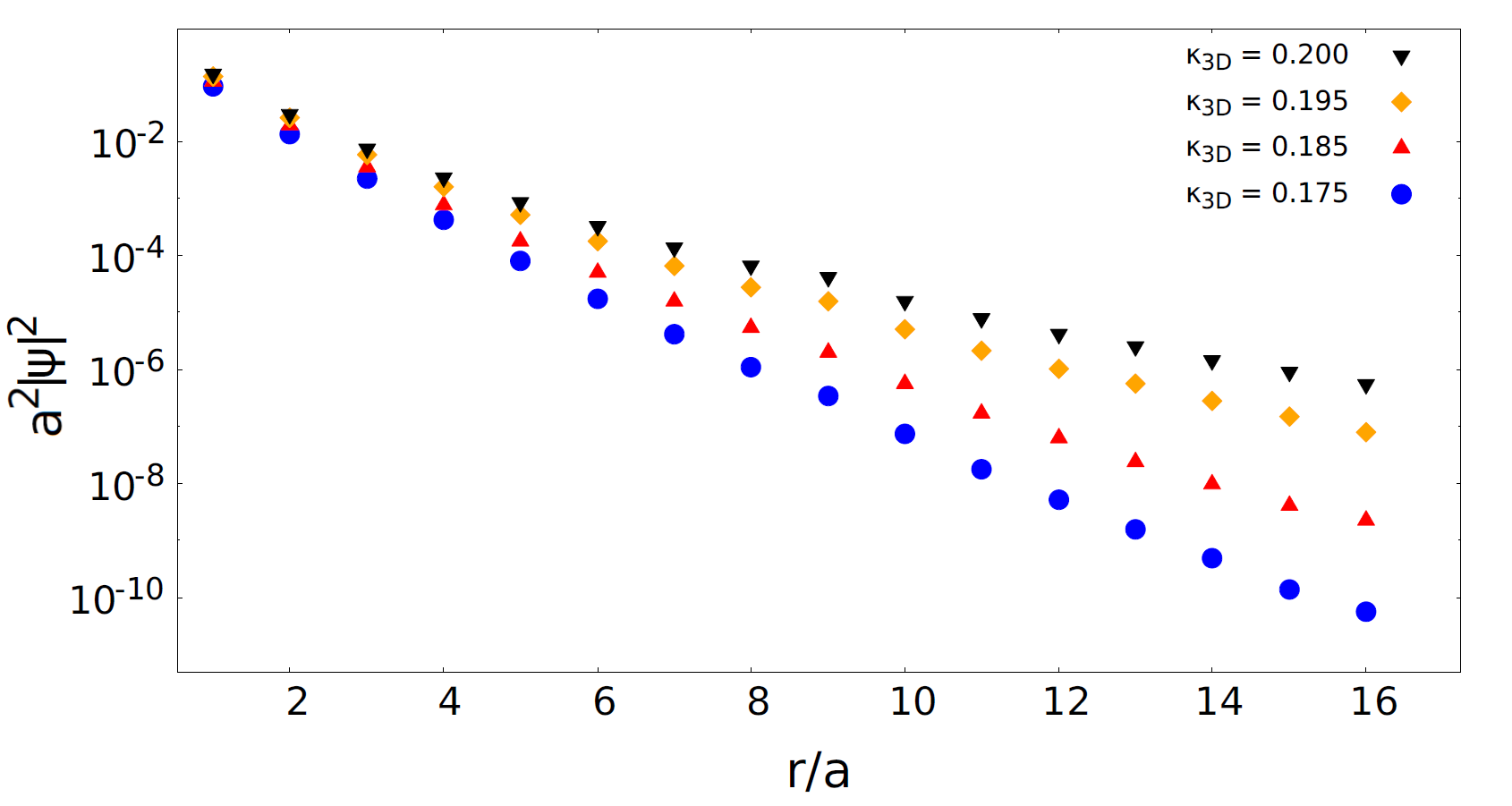}
	\caption{The shape of the modulus squared of the 3D propagator $\rvert\psi\rvert^2 = \rvert S_{3D}(0,r)\rvert^2$, the source of the extended operators, at different values of $\kappa_{3D}$. It is easy to see that the higher values, and thus the lighter 3D fermions, produce more extended sources (and sinks) along a spatial axis.}
	\label{fig:normkappas3dnormalized}
\end{figure}	
	\subsection{Parameters of the 3D operators}
\begin{figure}[h!]
	\centering
	\includegraphics[width=0.5\linewidth]{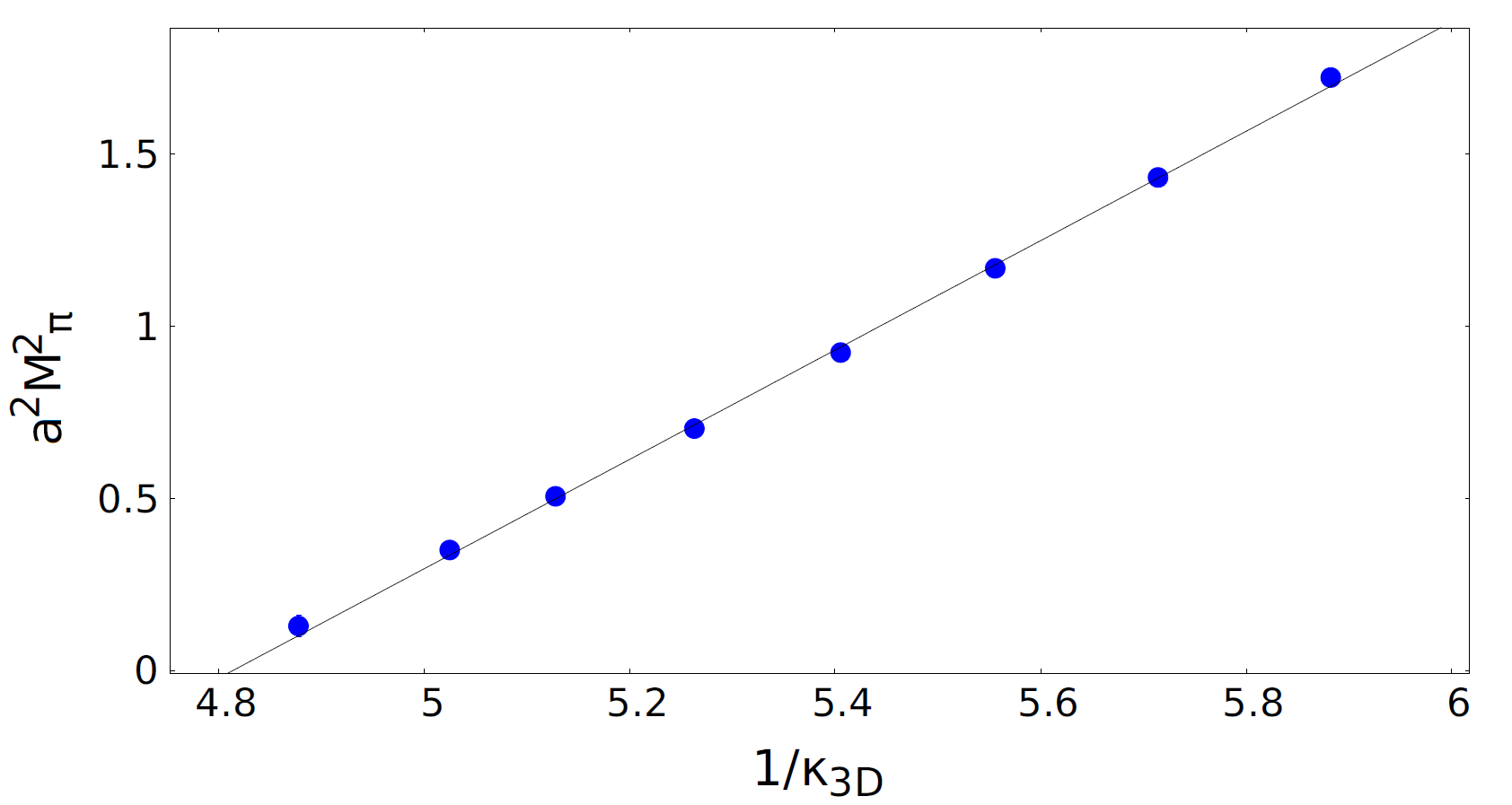}
	\caption{Different squared masses  of the 3D pion are calculated as a function of $\kappa_{3D}$. A linear fit is performed to extrapolate the critical value of $\kappa_{3D}$ for which the pion is massless.}
	\label{fig:m2pi}
\end{figure}

If we recall the action of the 3D fermions written in Eq. \ref{3daction}, we see that the 3D extended operators depend upon a set of masses of the 3D fermions.\\
Since this is an exploratory study we have imposed for both the 3D and ordinary quark fields an $SU(3)$ flavour symmetry. This means that the operators have only one free parameter, the 3D mass $ m_{3D}$. This is related to the 3D analogue of the $\kappa$ parameter as follows
\begin{equation}
am_{3D}  = \frac{1}{2}\left( \frac{1}{\kappa_{3D}}-6\right)
\end{equation}
$m_{3D}$ regulates the width of the extended operators. Heavier 3D fermions will result in more point-like operators while lighter ones will result in more extended objects as can be seen from Fig. \ref{fig:normkappas3dnormalized}, where the modulus squared of the 3D propagator is plotted in log scale for different values of $\kappa_{3D}$. Naturally this can be used to create operators which couple differently to different states of interest.\\
The modulus squared of the $3D$ propagator can be interpreted as the propagator for a pion, that propagates in two spatial dimensions plus a fictitious temporal direction which coincides with the third spatial direction on a fixed time-slice. And therefore the width of the sources can be linked to the mass of this object.\\
Moreover, analogously to the four dimensional case,  the 3D quark mass $m_{3D}$ is related to the square of the mass of the 3D pion $M_{\pi}^{3D}$. 
This fact can be used to find the critical value of $\kappa_{3D}$, namely the value for which the 3D quarks and pion are massless. In the free theory it would be $\kappa_{3D} = \frac{1}{6}$, in the full theory we expect quantum corrections to change this value.\\
We have built the propagator for a 3D pion and we have let it propagate by considering one of the three spatial directions  as "time" and the other two as space and then we have averaged over all the possible choices. We extracted the value of its mass for different $\kappa_{3D}$ and extrapolated to zero. A linear fit to this behaviour leads to an estimate of
\begin{equation}
\nonumber
	\kappa_{3D}^{crit} = 0.208\pm 0.004
\end{equation}
for this ensemble as it is shown in Fig.\ref{fig:m2pi}.
 For the rest of the
 study, we employ $\kappa_{3D} = 0.185$, which we have chosen to match
standard parameters of Jacobi sources as explained below.\\
The fact that the free parameters that define the $3D$ extended operators are quark masses means that these operators grant us theoretical control over the continuum limit. We know exactly how masses renormalize, moreover we can obtain the correct value of $m_{3D}$ for every value of the lattice spacing $a$ simply by matching the values of the 3D pion mass, which is a renormalization group invariant. \\
	\begin{figure}[h!]
	\centering
	\includegraphics[width=0.5\linewidth]{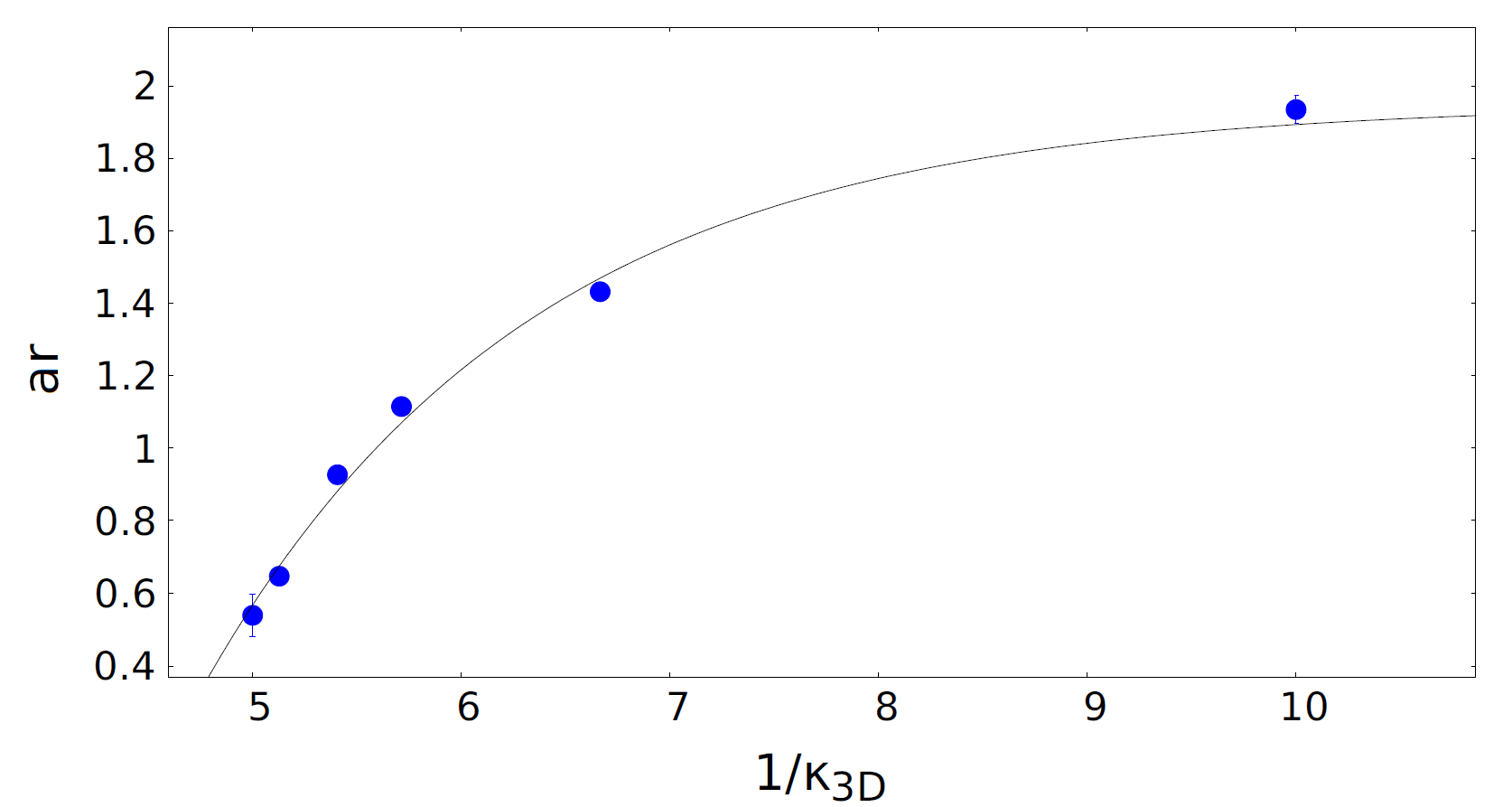}
	\caption{The strength to which the 3D extended operators couple to the excited states as a function of $1/\kappa_{3D}$. It is evident that the coupling decreases when approaching the critical value of $\kappa_{3D}$.}
	\label{fig:rvalue}
\end{figure}
We have mentioned that since the mass of the 3D fermions affects the shape of the 3D extended operators, it must also change how they couple to different states. To show this fact, we have computed the effective mass of the Nucleon and we have fitted it with the following function
\begin{equation}
\label{meff}
M_{eff}(x_0) = M + r e^{-(E-M) x_0}
\end{equation}
where we have modelled the coupling to the excited states in a single term with a parameter $r$ which regulates the strength of the coupling. We have plotted $r$ as a function of $\kappa_{3D}$ in Fig.\ref{fig:rvalue}, which shows that the coupling to the excited states changes rather dramatically by changing $\kappa_{3D}$. The more extended the operators are, the less they couple to the excited states.
\subsection{Results with the 3D extended operators}
\begin{figure}[h!]
	\centering
	\subfloat[3D3D]{\includegraphics[width=0.49\textwidth,height=\textheight,keepaspectratio]{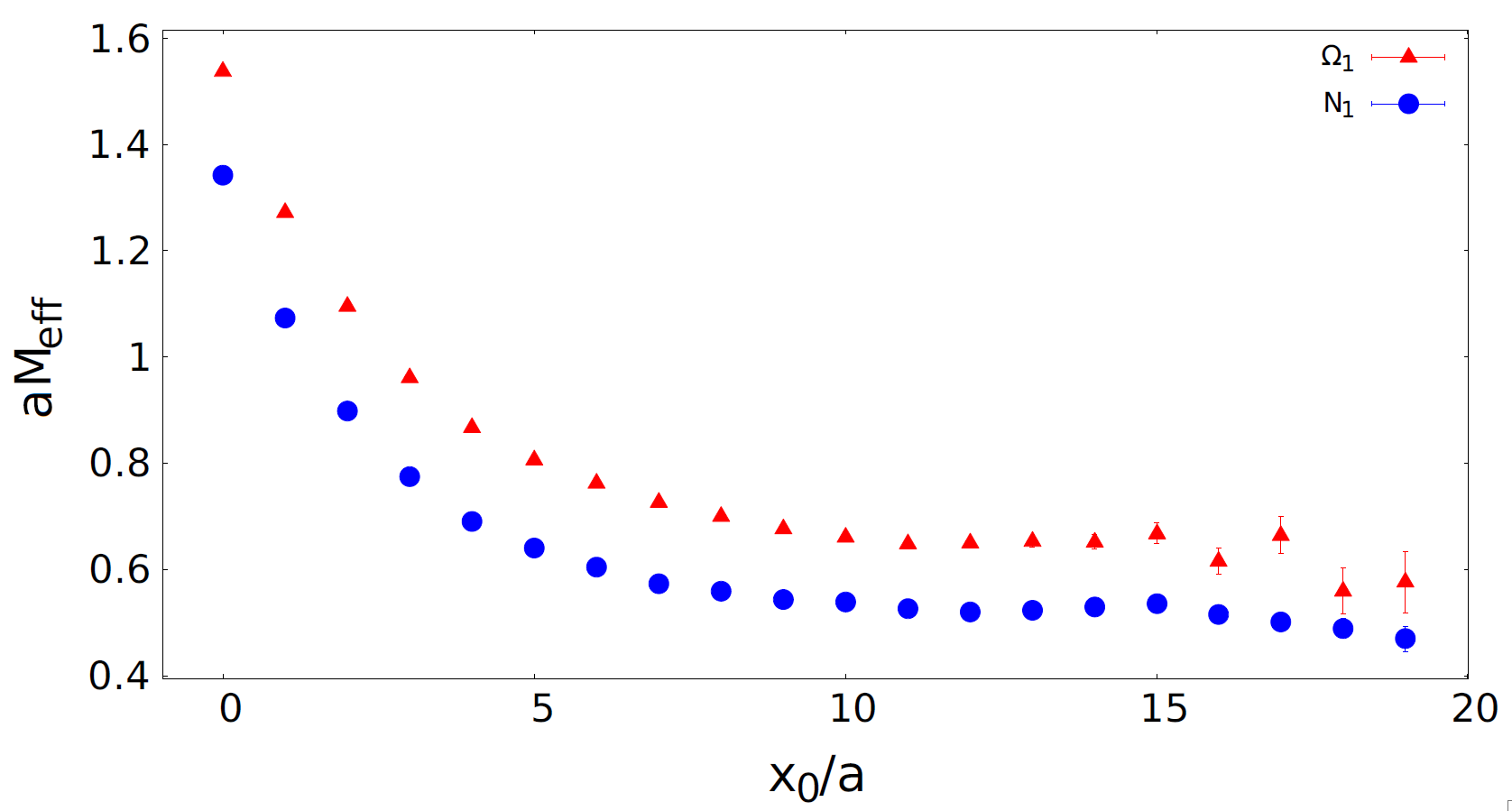}}
	\subfloat[3DP]{\includegraphics[width=0.49\textwidth]{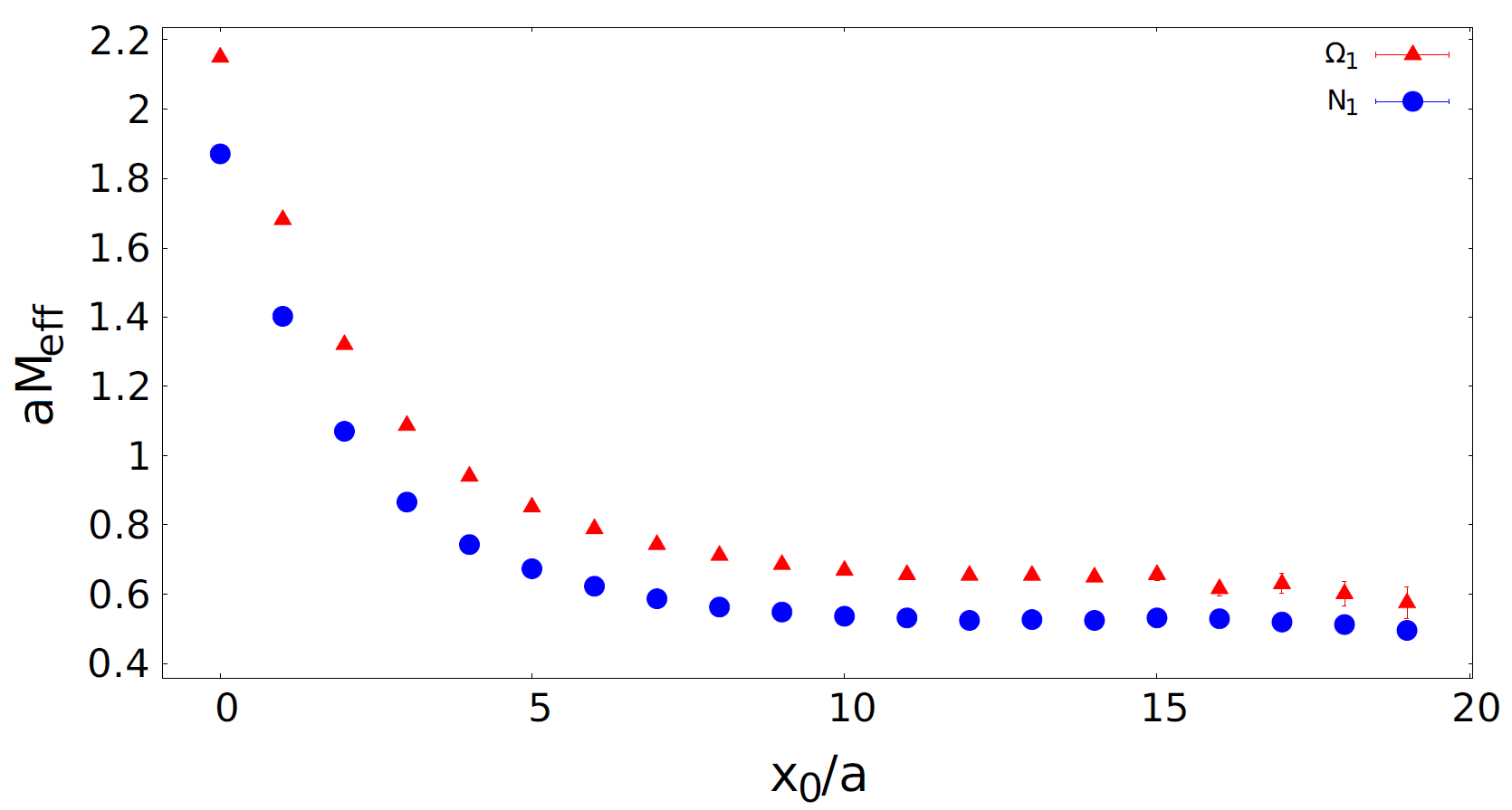}}
	\caption{The effective masses of the nucleon (circles) and of the Omega (triangles) calculated for the operators $N_1$ and $\Omega_1$. The first graph represent the effective masses calculated with the 3D operators both at source and sink, the second has a point operator at the sink while the source is 3D extended.}
	\label{fig:shortd}
\end{figure}
As mentioned in the introduction, we test the 3D extended operators by studying the two point functions of the nucleon and the Omega baryons, in particular we have obtained the correct interpolating operators as was described in section 2. This has given us a basis of interpolating operators for the spin $\frac{1}{2}$ octet nucleon $\{N_0,N_1,N_2\}$. These are the parity $P = +1$ eigenstates obtained from the operators in Eq. \ref{proton12} and \ref{proton3}. For the spin $\frac{3}{2}$ decouplet Omega we obtained the basis $\{\Omega_0,\Omega_1\}$, i.e. the parity $P = +1$ eigenstates obtained from the operators defined in Eq. \ref{omega12}. \\
As it is shown in Fig. \ref{fig:shortd}, the behaviour of the effective mass for the operators $N_1$ and $\Omega_1$ calculated with the 3D extended operators is very good, the short distance behaviour is regular and there is a clear plateau region. In the case of two extended 3D operators with the same 3D mass, both at source and sink, the value of the masses has been determined from a constant fit in the plateau region
\begin{align}
\nonumber
&a m_{N} =0.526\pm 0.002\\
\nonumber 
&a m_{\Omega} = 0.649 \pm 0.003 \ .
\end{align}
Note that, since we are working with configurations that have $M_\pi = M_K = 420\ MeV$ we expect the nucleon to be heavier and the Omega to be lighter then the real ones.
	\subsection{Comparison with Jacobi smearing}
\begin{figure}[H]
		\centering
		\subfloat[3D3D-JJ]{\includegraphics[width=0.49\textwidth,height=\textheight,keepaspectratio]{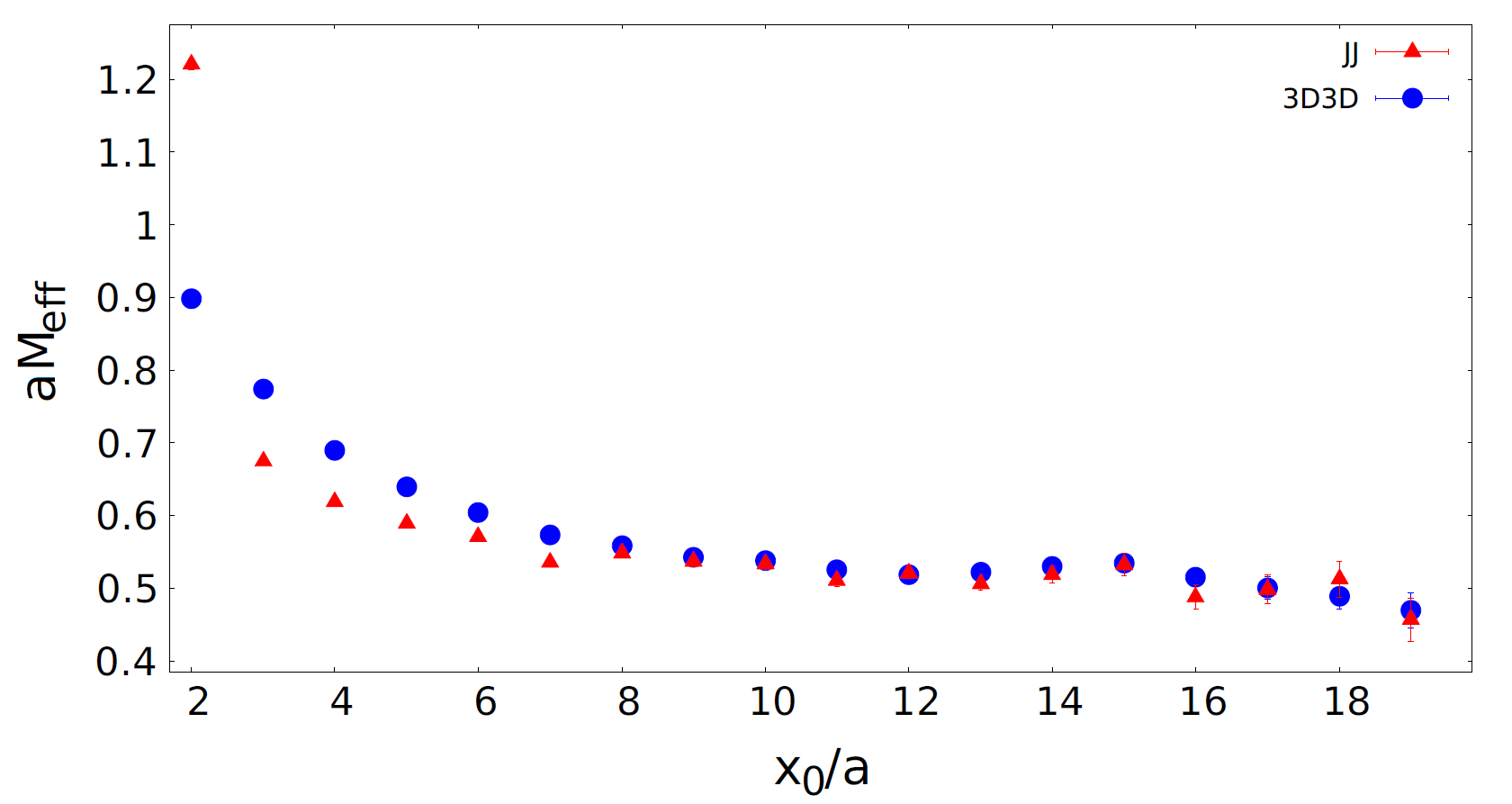}}
		\subfloat[3D3D-JP]{\includegraphics[width=0.49\textwidth]{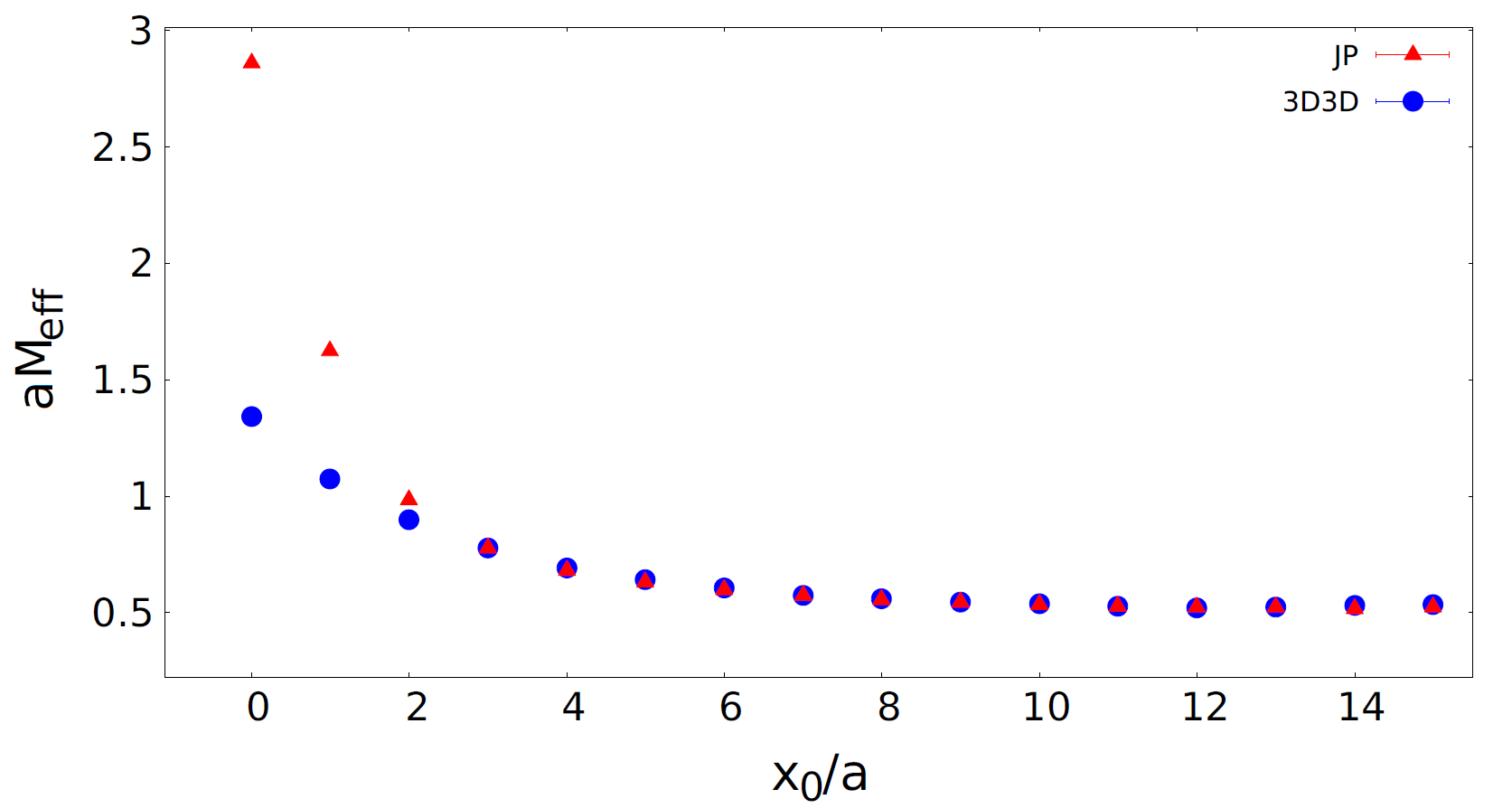}}
		\hfill
		\subfloat[3D3D-PP]{\includegraphics[width=0.49\textwidth,height=\textheight,keepaspectratio]{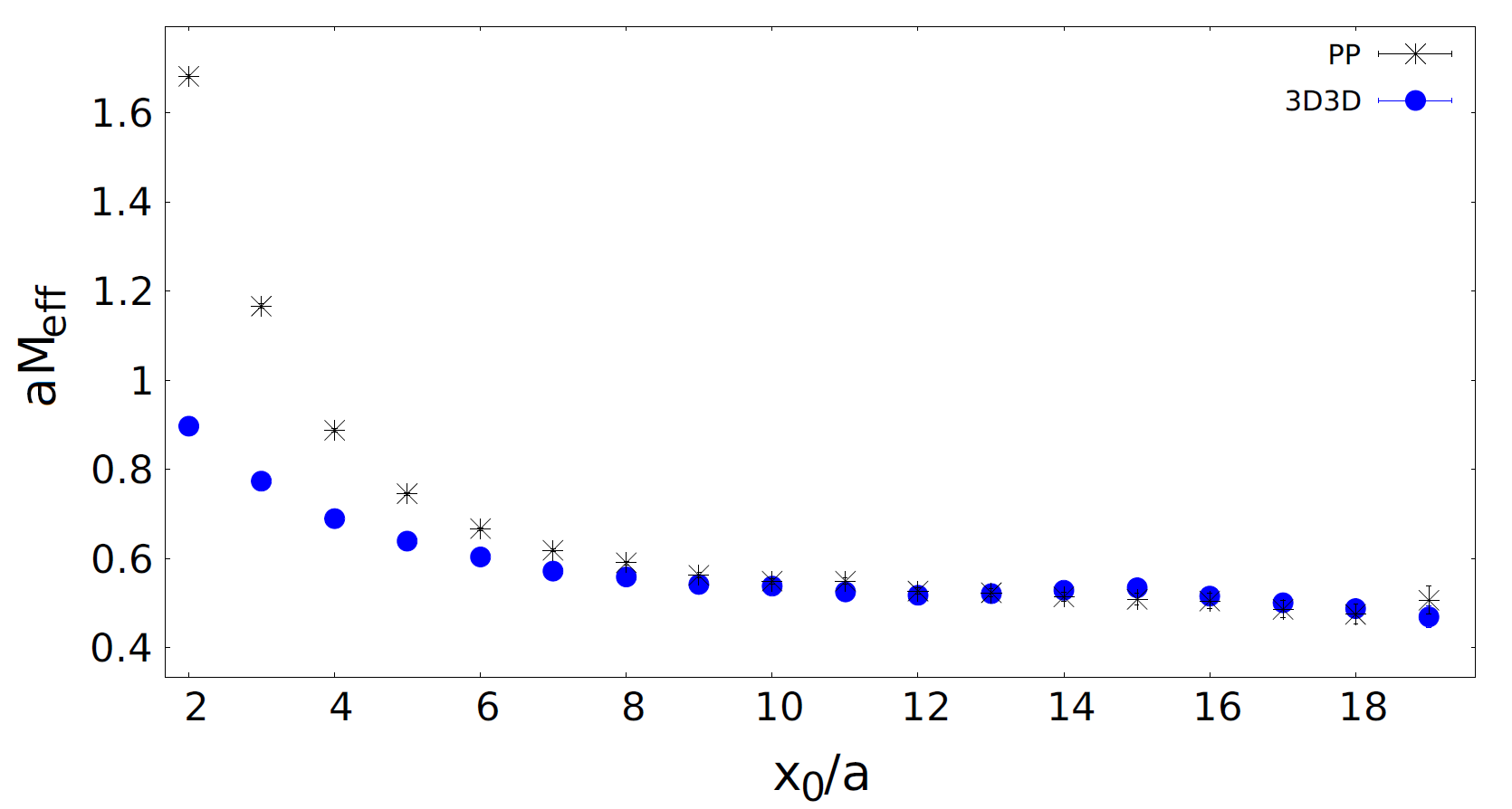}}
		\subfloat[3D3D-PP-JP]{\includegraphics[width=0.49\textwidth]{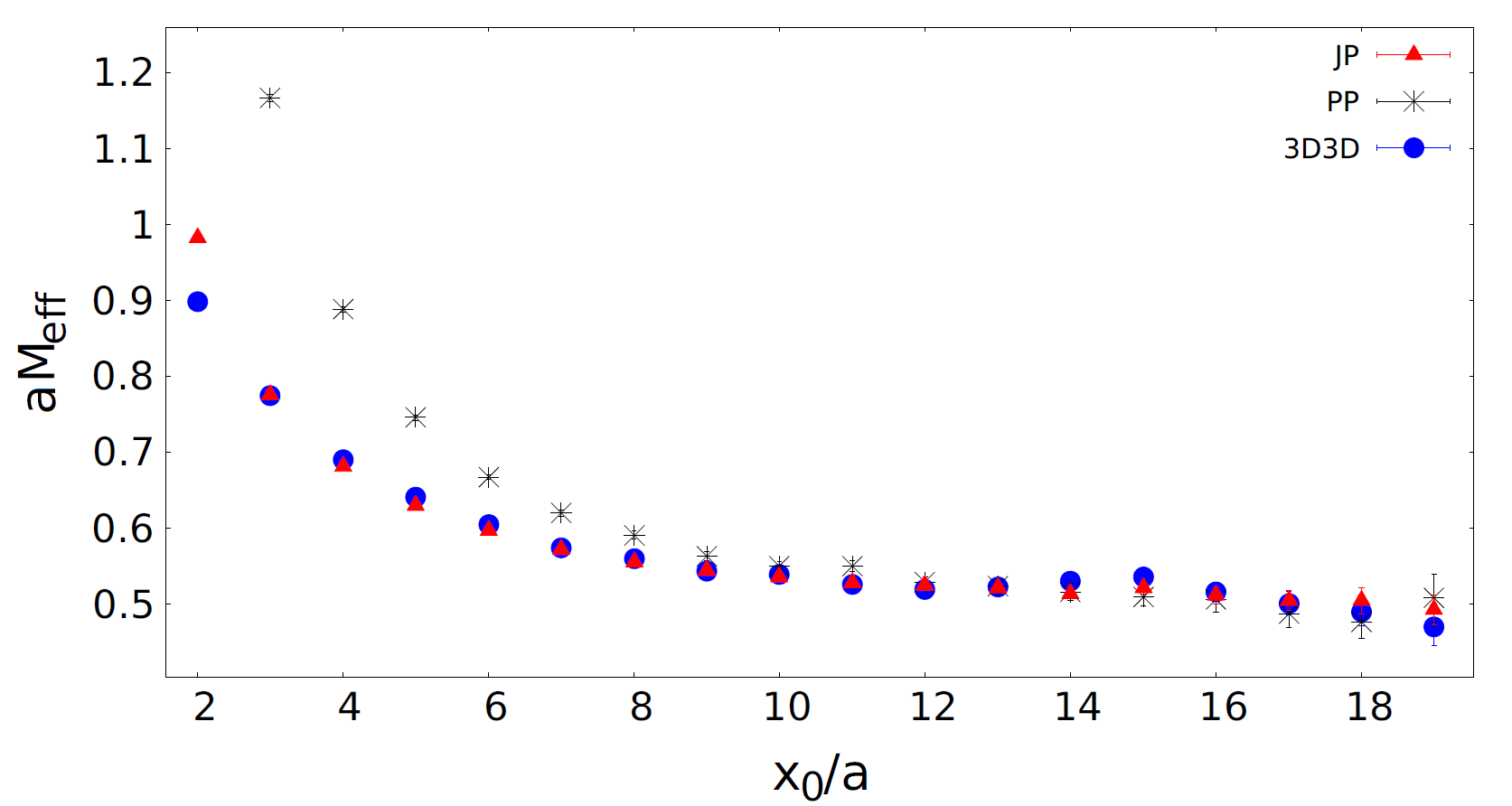}}
		\caption{The comparison between the effective masses of the nucleon calculated for different operators with an emphasis to the plateau region. It is possible to see that the Jacobi smearing and 3D sources produce very similar plateaus, it is also clear that the Jacobi smearing both in the sink and source (JJ) is much noisier then the 3D3D operator. The point sources are shown as a reference point, it is clear they cannot even produce a plateau.}
		\label{fig:plateau}
	\end{figure}

	To test the effectiveness of the 3D extended operator technique relative to the existing ones, we have computed the baryonic two point functions with  standard  point sources and Jacobi smeared sources.
	The free parameters of the Jacobi smearing are the number of terms included into the sum $N_{sm}$ and $\kappa_{sm}$ which regulates how strongly the source is spread out \cite{Allton:1993wc}  
	\begin{equation}
	\label{j_source}
	\Psi_\mathrm{sm} = \sum_{n = 0}^{N_{sm}} (\kappa_{sm}\Delta)^{n}\  \Psi_\mathrm{pnt} \,.
	\end{equation}
		\begin{figure}[h!]
		\centering
		\includegraphics[width=0.5\linewidth]{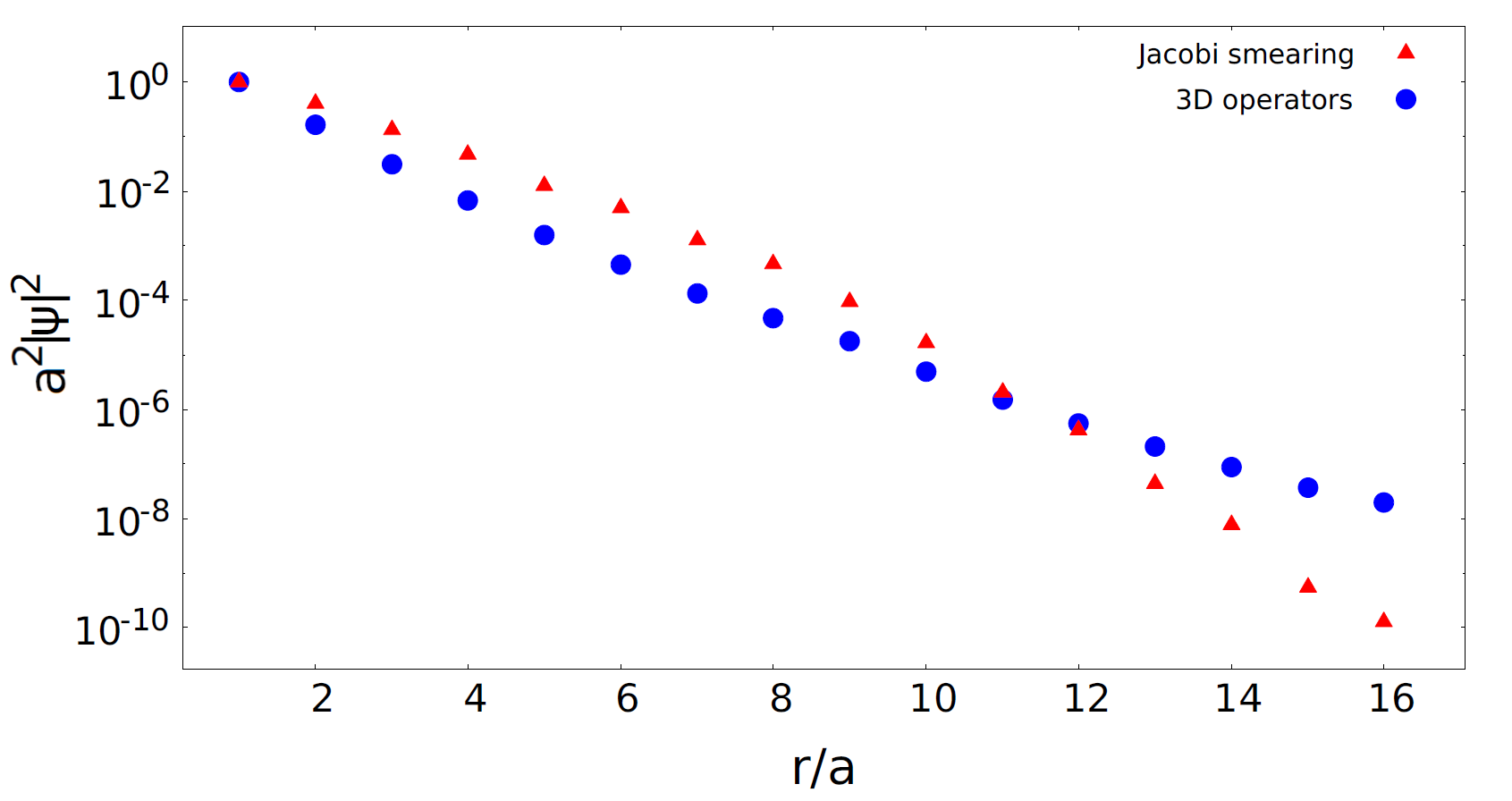}
		\caption{The norm of the 3D source $\rvert\psi\rvert^2 = \rvert S_{3D}(0,r)\rvert^2$ (circles)  and the Jacobi smeared source (triangles) defined in Eq. \ref{j_source} as a function of distance $r$ for the extended operators. The parameters of the Jacobi smearing were tuned to match the average radii $\langle r^2 \rangle$ along a spatial axis. }
		\label{fig:psisquarekappas2}
	\end{figure}
	The parameters have been chosen to be $N_{sm}  = 50$ and $\kappa_{sm} = 0.20$,
	such that the square of the source radius matches to the one of the 3D fermions.
	The Jacobi smeared source $\Psi_{sm}$ behaves very differently from
	the 3D one as can be seen in Fig.~\ref{fig:psisquarekappas2}. The latter has the typical exponential 
	decay in the long distance region, whereas the Jacobi source has a shape similar to a Gaussian. 
	The different shapes of these two source types give the opportunity to construct operators that differ
	significantly from one another.\\
	As can be seen from Fig. \ref{figure_shortd}, the two point functions obtained from the Jacobi and Point sources have a very singular behaviour at short distances while the 3D operators are much better behaved.
	Furthermore we can clearly infer from a zoom onto the plateau region, as it is shown in Fig. \ref{fig:plateau}, that Jacobi and 3D extended operators yield very similar results in terms of quality of the signal.\\
	The effective masses $M_{eff}(x_0)$ have been fitted with the same function we have defined in Eq.\ref{meff} in the range $ \left[t,t_{max}\right]$, this has allowed us to model the coupling to the excited states with a parameter $r$ and to study the stability of the different methods when they are studied as functions of $t$.\\
	The tables \ref{tab3d1} and \ref{tabj1} show the behaviour of the parameters of the fit as a function of the starting time $t$ for a single 3D and Jacobi operator respectively.\\ The single 3D operator is very stable when  starting the fit at very short distances and remains stable starting fit at larger times. On the other hand the fit of a single Jacobi operator is unstable at short distances and converges very slowly at larger times where the signal begins to fade into noise much faster that in the 3D case. So the quality of the 3D signal is not really equivalent to the one of the Jacobi sources.\\
 The comparison between the different methods also requires a discussion of
their computational cost. The  evaluation
of the two point function of the 3D operators requires two inversions of
the 3D
Dirac operator (at the sink and the source) and one inversion of the four
dimensional Dirac operator.\\
At our simulation parameters,  Jacobi smearing and 3D operators take roughly
the same computer time, which is not surprising since we were aiming at a
comparable smoothing radius for both methods. Also for the 3D sources,
smaller radii
lead to significantly reduced computer time needed to construct sources and
sinks. For the typical parameter $\kappa_{3D}=0.185$ this means that
constructing the source takes roughly $5\%$ of the time it takes to
invert the
4D Dirac operator. Even the 3D inversion on all time slices, needed at
the sink locations,
can be done in a fifth of the time of the latter.\\
Note that in this
comparison, the 4D operator is inverted using the advanced locally deflated
solver of the {\tt openQCD} code, whereas the equation involving the 3D
operator is
solved with a simple conjugate gradient algorithm. In case of the need
of even
wider sources, or many smoothing radii, the deflation techniques could
certainly also be used to significantly speed up the 3D inversions.
\renewcommand{\arraystretch}{1.5}
\begin{table}[ht]
	\centering
	\begin{adjustbox}{width=1.1\textwidth,center=\textwidth}
		\begin{tabular}{ | p{1.5cm} | p{3cm} | p{3cm}  | p{3cm}| p{1cm}|}
			\hline
			Basis & $ar$ & $am$  & $aE_1$ &$t/a$\\ 
			\hline                       
			$3D_1$ & 0.820 $\pm$ 0.003 & 0.519 $\pm$ 0.0024  & 0.907 $\pm$ 0.006 & 1\\
			$3D_1$ & 0.811 $\pm$ 0.007 & 0.517 $\pm$ 0.0027  & 0.898 $\pm$ 0.009 & 2\\
			$3D_1$ & 0.811 $\pm$ 0.017 & 0.517 $\pm$ 0.0032  & 0.898 $\pm$ 0.013 & 3\\
			$3D_1$ & 0.767 $\pm$ 0.033 & 0.518 $\pm$ 0.0037  & 0.878 $\pm$ 0.019 & 4\\
			$3D_1$ & 0.692 $\pm$ 0.057 & 0.512 $\pm$ 0.0044  & 0.849 $\pm$ 0.027 & 5\\
			$3D_1$ & 0.733 $\pm$ 0.127 & 0.513 $\pm$ 0.0051  & 0.863 $\pm$ 0.042 & 6\\
			$3D_1$ & 0.808 $\pm$ 0.281 & 0.514 $\pm$ 0.0059  & 0.880 $\pm$ 0.069 & 7\\
			$3D_1$ & 0.504 $\pm$ 0.307 & 0.509 $\pm$ 0.0093  & 0.805 $\pm$ 0.109 & 8\\
			$3D_1$ & 0.640 $\pm$ 0.807 & 0.511 $\pm$ 0.0106  & 0.839 $\pm$ 0.186 & 9\\
			\hline
		\end{tabular}
	\end{adjustbox}
	\caption{the values of $r$,$m$ and $E_1$ for the $3D_1$ operator for different starting point of the fit $t$ (increasing number of excluded points). It is possible to appreciate the remarkable stability of the fit and its low uncertainties. The table points stop when the fit becomes unstable.} 
	\label{tab3d1}
\end{table}
\renewcommand{\arraystretch}{1.5}
\begin{table}[ht]
	\centering
	\begin{adjustbox}{width=1.1\textwidth,center=\textwidth}
		\begin{tabular}{ | p{1.5cm} | p{3cm} | p{3cm}  | p{3cm}| p{1cm}|}
			\hline
			Basis & $ar$ & $am$  & $aE_1$ &$t/ a$\\ 
			\hline     
			$J_1$ & 4.828 $\pm$ 0.329  & 0.456 $\pm$ 0.0959 & 1.159 $\pm$ 0.169 & 1\\
			$J_1$ & 16.981$\pm$ 0.685  & 0.548 $\pm$ 0.0074 & 2.157 $\pm$ 0.042 & 2\\
			$J_1$ & 11.871$\pm$ 3.386  & 0.544 $\pm$ 0.0081 & 1.981 $\pm$ 0.143 & 3\\
			$J_1$ & 0.479 $\pm$ 0.068  & 0.512 $\pm$ 0.0063 & 0.877 $\pm$ 0.052 & 4\\
			$J_1$ & 0.384 $\pm$ 0.093  & 0.508 $\pm$ 0.0086 & 0.819 $\pm$ 0.075 & 5\\
			$J_1$ & 0.357 $\pm$ 0.161  & 0.506 $\pm$ 0.0110 & 0.803 $\pm$ 0.113 & 6\\
			$J_1$ & 0.213 $\pm$ 0.125  & 0.497 $\pm$ 0.0238 & 0.692 $\pm$ 0.163 & 7\\
			$J_1$ & 0.382 $\pm$ 2.062  & 0.194 $\pm$ 2.106  & 0.208 $\pm$ 2.201 & 8\\
			\hline
		\end{tabular}
	\end{adjustbox}
	\caption{the values of $r$,$m$ and $E_1$ for the $J_1$ operator for different starting point of the fit $t$ (increasing number of excluded points). The fit takes time to converge and by the time it has stabilized the noise start to overtake the signal.  The table points stop when the fit becomes unstable.} 
	\label{tabj1}
\end{table}
\FloatBarrier
\subsubsection{GEVP}
The 3D operators can be used to build or enlarge a basis of linearly independent operators that can be used to perform an analysis based on variational methods like the GEVP.\\
The GEVP method consist in building an hermitian matrix of correlators from a basis of linearly independent operators
\begin{equation}
C_{ij}(x_0) = \int d^3\mathbf{x}\ \langle O_i \bar{O}_j \rangle
\end{equation}
and then to solve  the corresponding generalized eigenvalue problem
\begin{equation}
C(x_0) v_n = \lambda_n(x_0,t_0) C(t_0) v_n,\quad n = 1,\ldots,N \ .
\end{equation}
It has been shown \cite{Luscher:1990ck} \cite{Blossier:2009kd} that the eigenvalues $\lambda_n(x_0,t_0)$ in an ideal situation correspond to $N$ energy states interpolated by the set of linearly independent operators $O_i$ via
\begin{equation}
	E_n = \log \frac{\lambda_n(x_0,t_0)}{\lambda_n(x_0+a,t_0)}
\end{equation}
where $E_0 = M_{eff}$ is the effective mass of the ground state.\\
We have verified that the 3D extended operators are stable for any value of $t_0$, having very small short distance divergences, on the other hand the Jacobi and point sources need a larger $t_0$ to move away from the divergent short distance behaviour, in the analysis that follows $t_0 = 4 a$ to allow for a fair comparison between methods.\\
The GEVP has been performed on a $2\times2$ basis corresponding to the $\{N_0,N_1\}$ operators.
As can be seen from tables \ref{tab3d}, \ref{tabj} and \ref{tabp} the $3D$ operators are remarkably stable even at short distances, the parameters of the fit remain constant, on the other hand Jacobi and point sources need time to stabilize and are overall less reliable. In particular the point sources have their parameters fluctuating much more than their uncertainties range.\\
We can also compare how the GEVP  fares in comparison with the single operators we have studied previously. By confronting tables, it is clear that the GEVP does not improve the results obtained with a single 3D operator. On the other hand the GEVP  on the Jacobi smeared sources  show a small improvement in respect to a single source.
	\renewcommand{\arraystretch}{1.5}
	\begin{table}[ht]
		\centering
		\begin{adjustbox}{width=1.1\textwidth,center=\textwidth}
			\begin{tabular}{ | p{2.5cm} | p{3cm} | p{3cm}  | p{3cm}| p{1cm}|}
				\hline
				Basis & $ar$ & $am$  & $aE_1$ &$t/a$\\ 
				\hline                       
				$\{3D_0,3D_1\}$ & 0.818 $\pm$ 0.003 & 0.521 $\pm$ 0.0025  & 0.912 $\pm$ 0.006 & 1\\
				$\{3D_0,3D_1\}$ & 0.813 $\pm$ 0.006 & 0.519 $\pm$ 0.0028  & 0.906 $\pm$ 0.008 & 2\\
				$\{3D_0,3D_1\}$ & 0.820 $\pm$ 0.016 & 0.520 $\pm$ 0.0033  & 0.910 $\pm$ 0.013 & 3\\
				$\{3D_0,3D_1\}$ & 0.776 $\pm$ 0.034 & 0.518 $\pm$ 0.0038  & 0.891 $\pm$ 0.019 & 4\\
				$\{3D_0,3D_1\}$ & 0.726 $\pm$ 0.065 & 0.516 $\pm$ 0.0046  & 0.873 $\pm$ 0.028 & 5\\
				$\{3D_0,3D_1\}$ & 0.841 $\pm$ 0.158 & 0.518 $\pm$ 0.0050  & 0.905 $\pm$ 0.045 & 6\\
				$\{3D_0,3D_1\}$ & 1.168 $\pm$ 0.489 & 0.521 $\pm$ 0.0050  & 0.961 $\pm$ 0.078 & 7\\
				$\{3D_0,3D_1\}$ & 0.876 $\pm$ 0.745 & 0.519 $\pm$ 0.0068  & 0.918 $\pm$ 0.137 & 8\\
				$\{3D_0,3D_1\}$ & 0.388 $\pm$ 0.586 & 0.515 $\pm$ 0.0131  & 0.804 $\pm$ 0.231 & 9\\
				\hline
			\end{tabular}
		\end{adjustbox}
		\caption{the values of $r$,$m$ and $E_1$ for the $\{3D_0,3D_1\}$ basis for $t_0 = 4a$ for different starting point of the fit $t$ (increasing number of excluded points).  The table points stop when the fit becomes unstable.} 
						\label{tab3d}
	\end{table}
\begin{table}[ht]
	\centering
	\begin{adjustbox}{width=1.1\textwidth,center=\textwidth}
		\begin{tabular}{ | p{2.5cm} | p{3cm} | p{3cm}  | p{3cm}| p{1cm}|}
			\hline
			Basis & $ar$ & $am$  & $aE_1$ &$t/a$\\   
			\hline  
			$\{J_0,J_1\}$ & 4.843  $\pm$ 0.339  & 0.457 $\pm$ 0.0932  & 1.164 $\pm$ 0.166 & 1\\
			$\{J_0,J_1\}$ & 17.469 $\pm$ 0.662  & 0.540 $\pm$ 0.0065  & 2.162 $\pm$ 0.039 & 2\\
			$\{J_0,J_1\}$ & 17.127 $\pm$ 5.512  & 0.539 $\pm$ 0.0072  & 2.152 $\pm$ 0.161 & 3\\
			$\{J_0,J_1\}$ & 0.403 $\pm$ 0.068  & 0.511 $\pm$ 0.0066  & 0.865 $\pm$ 0.061 & 4\\
			$\{J_0,J_1\}$ & 0.298 $\pm$ 0.080  & 0.505 $\pm$ 0.0097  & 0.786 $\pm$ 0.084 & 5\\
			$\{J_0,J_1\}$ & 0.274 $\pm$ 0.131  & 0.503 $\pm$ 0.0127  & 0.766 $\pm$ 0.124 & 6\\
			$\{J_0,J_1\}$ & 0.157 $\pm$ 0.057  & 0.485 $\pm$ 0.0424  & 0.619 $\pm$ 0.180 & 7\\
			$\{J_0,J_1\}$ & 0.428 $\pm$ 2.898  & 0.141 $\pm$ 2.942  & 0.151 $\pm$ 3.032 & 8\\
			\hline
		\end{tabular}
	\end{adjustbox}
	\caption{the values of $r$,$m$ and $E_1$ for the $\{J_0,J_1\}$ basis for $t_0 = 4a$ for different starting point of the fit $t$ (increasing number of excluded points). The fit makes sense until $t = 7a$ the last point is included to illustrate this fact. } 
				\label{tabj}
\end{table}
\renewcommand{\arraystretch}{1.5}
\begin{table}[ht]
	\centering
	\begin{adjustbox}{width=1.1\textwidth,center=\textwidth}
		\begin{tabular}{ | p{2.5cm} | p{3cm} | p{3cm}  | p{3cm}| p{1cm}|}
			\hline
			Basis & $ar$ & $am$  & $aE_1$ &$t/a$\\ 
			\hline      
			$\{P_0,P_1\}$ & 2.628 $\pm$ 0.060  & 0.459 $\pm$ 0.0445 & 0.874 $\pm$ 0.068 & 1\\
			$\{P_0,P_1\}$ & 3.366 $\pm$ 0.047  & 0.532 $\pm$ 0.0081 & 1.102 $\pm$ 0.018 & 2\\
			$\{P_0,P_1\}$ & 3.877 $\pm$ 0.138  & 0.544 $\pm$ 0.0060 & 1.167 $\pm$ 0.020 & 3\\
			$\{P_0,P_1\}$ & 3.141 $\pm$ 0.219  & 0.535 $\pm$ 0.0058 & 1.097 $\pm$ 0.026 & 4\\
			$\{P_0,P_1\}$ & 2.104 $\pm$ 0.177  & 0.523 $\pm$ 0.0048 & 0.989 $\pm$ 0.026 & 5\\
			$\{P_0,P_1\}$ & 1.381 $\pm$ 0.159  & 0.514 $\pm$ 0.0049 & 0.896 $\pm$ 0.029 & 6\\
			$\{P_0,P_1\}$ & 0.980 $\pm$ 0.183  & 0.507 $\pm$ 0.0065 & 0.830 $\pm$ 0.041 & 7\\
			$\{P_0,P_1\}$ & 0.647 $\pm$ 0.188  & 0.498 $\pm$ 0.0103 & 0.755 $\pm$ 0.061 & 8\\
			$\{P_0,P_1\}$ & 0.365 $\pm$ 0.122  & 0.478 $\pm$ 0.0257 & 0.640 $\pm$ 0.095 & 9\\
			\hline
		\end{tabular}
	\end{adjustbox}
	\caption{the values of $r$,$m$ and $E_1$ for the $\{P_0,P_1\}$ basis for $t_0 = 4a$ for different starting point of the fit $t$ (increasing number of excluded points). The parameters fluctuate well beyond their uncertainty ranges. The table points stop when the fit becomes unstable.} 
	\label{tabp}
\end{table}
\FloatBarrier
\section{Conclusions}
The quenched 3D fermions that we have introduced allow for a formulation of the extended operators in terms of a renormalizable field theory. The only free parameter of this theory is the mass (or the masses) of the 3D fermions which is easily adjustable through its connection with the pseudoscalar meson mass. This framework gives theoretical control over the 3D extended operators, which can be considered as a viable alternative to the standard smearing procedures.
In the present work we have shown that the 3D extended operators improve the short distance behaviour of the
two point functions and that they are well behaved under renormalization so that the continuum limit can be taken in a controlled way. The disappearance of short distance divergences might
improve the study of excited states, where a strong signal is needed at short
Euclidean times.  However a satisfactory understanding of the  interplay
between the short distance behaviour and excited states can only be achieved
through a scaling study.  \\
In the numerical application, we have shown that the 3D extended operators are as computationally efficient as the Jacobi smearing for a large range of $3D$ masses even without optimizations for the evaluation of the 3D propagators. In fact these can be computed with standard 
iterative techniques for sparse systems, making them a computationally
efficient choice for the construction of wide sources. \\
We have shown through numerical simulations of the two point functions of the nucleon and the Omega 
that the 3D extended operators can be used to obtain very precise results.
The differences in the short distance behaviour with respect to Jacobi smearing
make the 3D extended operators a potentially very interesting complement to the basis used
in the GEVP and therefore they look promising in increasing the precision in the extraction of the baryon
spectrum. 

\section*{Acknowledgements}
M.P. is indebted to Martin  L\"uscher for many useful discussions. He also acknowledges partial support by the MIUR-PRIN grant 2010YJ2NYW and by the INFN SUMA project.
\newpage
\section*{Appendix A}
\begin{figure}
	\centering
	\includegraphics[width=1\linewidth]{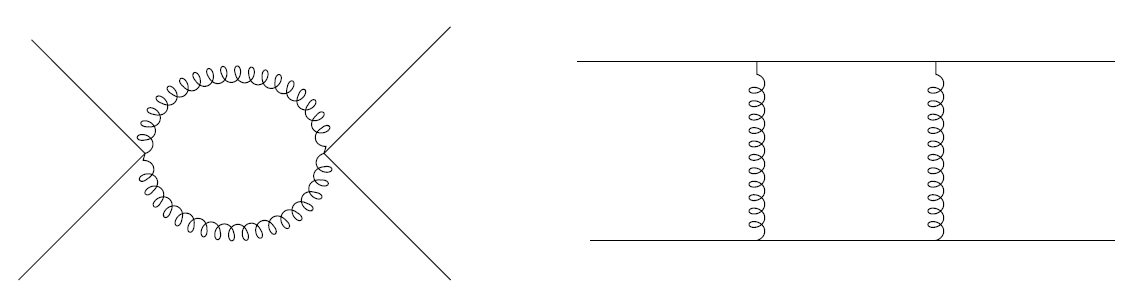}
	\caption{The two types of one loop diagrams associated with the four point functions of the 3D bosons generated by the interactions with the gauge field.}
	\label{fig:1loopdiag}
\end{figure}

In this appendix we show that a theory of quenched 3D bosonic fields coupled to the spatial components of the gauge field is inconsistent. Therefore, even though it would constitute a viable smearing technique at finite lattice spacing, it would not retain any of the advantages of a field theory formulation.\\
A theory of minimally coupled, massive, 3D bosons $\phi^i$ with $i = 1,\ldots,N$ to a $SU(N)$ 4D gauge field $A_\mu^a$ with $a = 1,\ldots,N$ can be described by the Lagrangian density
\begin{equation}
\mathcal{L}_B = \delta(x_0-\tau)\left[ \frac{1}{2}\left(D_\mu^{ij}\phi^j\right)^{\dagger}\left(D_\mu^{ik}\phi^k\right)-\frac{m^2_B}{2}\phi^{\dagger}_i\phi_i-\frac{\lambda_1}{4}\left(\phi^{\dagger}_i\phi_i\right)^2-\frac{\lambda_2}{6}\left(\phi^{\dagger}_i\phi_i\right)^3\right]
\end{equation}
where  $D^{ij}_\mu = \delta^{ij}\partial_\mu-i g A^{a}_\mu T_a^{ij}$ is the covariant derivative, $g$ is the dimensionless gauge coupling and $T^{a}_{ij}$ are the generators of the Lie algebra. Therefore the 3D boson fields $\phi^i$ have a classical canonical dimension of $\frac{1}{2}$. The 3D bosonic fields also obey the following constraint
\begin{equation}
	D_0^{ij}\phi^j \big\rvert_{x_0=\tau}= 0
\end{equation} 
so that they don't propagate in time. \\
The quartic and sexctic terms are present due to being, respectively, the only relevant $[\lambda_1] = 1$ and marginal $[\lambda_2] =0$ operators allowed by the symmetries of the system.
The action can be written as
\begin{equation}
S_B = \int d^3\mathbf{x}\ \frac{1}{2}\left(D_\mu^{ij}\phi^j\right)^{\dagger}\left(D_\mu^{ik}\phi^k\right)-\frac{m^2_B}{2}\phi^{\dagger}_i\phi_i-\frac{\lambda_1}{4}\left(\phi^{\dagger}_i\phi_i\right)^2-\frac{\lambda_2}{6}\left(\phi^{\dagger}_i\phi_i\right)^3
\end{equation}
If we expand the kinetic term of the action we find
\begin{equation}
\frac{1}{2}\rvert\partial_\mu\phi\rvert^2+\frac{i}{2}g A_\mu^a T^a_{ij}\left(\phi^\dagger_i \partial^\mu\phi_j-\phi_i \partial^\mu\phi^\dagger_j\right)+\frac{g^2}{2}A_\mu^aA^\mu_b T^a_{ij} T^b_{ik}\phi^j\phi^k
\end{equation}
so we have three interaction vertices.\\
At one loop the two point function of the $\phi$ fields will have divergent contributions
that renormalize the mass and the field. Furthermore, the four point function will also receive quantum corrections from the gauge fields of the type represented in Fig.\ref{fig:1loopdiag}. These objects, at one loop, will give divergent contributions of the type $\sim \phi^4$.\\
 This exposes the problem: the 3D bosons are quenched in order to be useful as a smearing technique. By quenched we mean that they are classical, as in the background field method $\phi=\phi_c+\delta \phi$ where we keep only the classical field $\phi_c$. If this was not the case, the quantum contributions of the 3D theory would change the four dimensional gauge theory.\\
However, the fact that we only have a classical 3D boson field means that the counterterms that come from the self interactions (in particular $\phi^4$ at one loop ) that should be generated to make finite the diagrams of Fig.\ref{fig:1loopdiag},  are not present and so the quenched theory is not renormalizable. In practice this means that, although the $3D$ bosons are used as a smearing technique on the lattice, the lack of renormalizability make them impossible to control from a theoretical point of view.\\
The fermionic 3D fields clearly do not suffer from this problem, in fact the 3D fermions $\varphi$ have canonical dimension of $1$ and therefore beyond a mass term $\bar{\varphi}\varphi$ which has dimension 2 (and so $[m_{3D}] = 1$)  there are no other relevant o marginal operators. For instance, an operator like $\lambda \left(\bar{\varphi}\varphi\right)^2$ has dimension $4$ and so $[\lambda]=-1$ making it irrelevant. This means that the diagrams of the 3D fermionic theory can be made finite even in the quenched theory.		
\section*{Appendix B}
We calculate the mass and field renormalization at one loop for the 3D fermion fields in the continuum to show that they have the same divergences as their four dimensional counterparts.
The 3D fermions two point function is defined as
\begin{equation}
S_{3D}(\mathbf{x},\mathbf{y}) = \langle \varphi(\mathbf{x})\bar{\varphi}(\mathbf{y})\rangle
\end{equation}
In the path integral the interaction vertex is of the form
\begin{equation}
V = g \int d^3\mathbf{x}\ \bar{\varphi} \gamma_i A^i_a T^a \varphi,\qquad i = 1,2,3
\end{equation}
where $a$ is the colour index while $i$ are the spatial Lorentz indices, notice that the gauge field is calculated on a fixed time-slice.\\
At one loop we will have the Wick contractions due to the contribution 
\begin{equation}
S^{1\ loop}_{3D}(\mathbf{x},\mathbf{y}) = \langle \varphi(\mathbf{x})\gamma_i V^i_a T^a \gamma_j V^j_b T^b\bar{\varphi}(\mathbf{y})\rangle 
\end{equation}
this can be written as 
\begin{align}
	S^{1\ loop}_{3D}(\mathbf{x},\mathbf{y}) =  \frac{ g^2 \Tr T^a T^a}{(2 \pi)^{13}}&\int d^3\mathbf{p}\ d^4q\ d^3\mathbf{k}\ d^3\mathbf{l}\ d^3\mathbf{u}\ d^3\mathbf{w}\ \frac{e^{i \mathbf{p(x-u)}}}{\slashed{\mathbf{p}}+m_{3D}}\\\nonumber
&	\times\gamma_i\frac{e^{i \mathbf{q(u-w)}}g^{ij}}{q^2}\frac{e^{i \mathbf{k(u-w)}}}{\slashed{\mathbf{k}}+m_{3D}}\gamma_j\frac{e^{i \mathbf{l(w-y)}}}{\slashed{\mathbf{l}}+m_{3D}}
\end{align}
where $u_0 = w_0$ since the vertices are on the same time-slice. The integrals in $\mathbf{w}$ and $\mathbf{u}$ give two Dirac deltas: $\delta^3(\mathbf{p-q-k})$ and $\delta^3(\mathbf{l-q-k})$, so one cancels the integral in $d^3\mathbf{k}$ and the other the one in $d^3\mathbf{l}$ fixing the external momenta to be $\mathbf{p} = \mathbf{l}$ as it should be. We are left with
\begin{align}
S^{1\ loop}_{3D}(\mathbf{x},\mathbf{y}) = \frac{ g^2 \Tr T^a T^a}{(2 \pi)^{7}}\int d^3\mathbf{p}\ d^4q \frac{e^{i\mathbf{ {p}(x-y)}}}{\slashed{\mathbf{p}}+m_{3D}}\gamma^i\frac{1}{q^2}\frac{1}{(\slashed{\mathbf{p}}-\slashed{\mathbf{q}})+m_{3D}}\gamma_i\frac{1}{\slashed{\mathbf{p}}+m_{3D}}
\end{align}
so we have the external legs and the loop integral inside which is the emission and re-absorption of a four dimensional gluon by the three dimensional fermions. 
The quantity between the external legs is the one loop contribution to the self energy
\begin{equation}
\Sigma^{1 loop}(\slashed{\mathbf{p}}) =  -\frac{ g^2 \Tr T^a T^a}{(2 \pi)^{4}}\int d^4q \frac{1}{q^2}\frac{(\slashed{\mathbf{p}}-\slashed{\mathbf{q}})-3 m_{3D}}{({\mathbf{p}}-{\mathbf{q}})^2+m^2_{3D}}
\end{equation}
we put this integral in a slightly more convenient form using the Feynman change of variables
\begin{equation}
\Sigma^{1 loop}(\slashed{\mathbf{p}}) = - \frac{ g^2 \Tr T^a T^a}{(2 \pi)^{4}} \int_{0}^{1}dx\ \int d^4q\ \frac{(\slashed{\mathbf{p}}-\slashed{\mathbf{q}})-3m_{3D}}{\left[({\mathbf{p}}-{\mathbf{q}})^2(1-x) +m^2_{3D}(1-x)+q^2x\right]^2}
\end{equation}
since we want to prove that the integral proportional to $\slashed{\mathbf{q}}$ is zero. The denominator can be rewritten as
\begin{equation}
(\mathbf{q}-\mathbf{p}(1-x))^2+\mathbf{p}^2 x(1-x)+m^2_{3D}(1-x)+q_0^2 x
\end{equation}
so now we can simply change variable to $\mathbf{q}\rightarrow\mathbf{q}-\mathbf{p}(1-x)$ and we get
\begin{equation}
\Sigma^{1 loop}(\slashed{\mathbf{p}}) =  -\frac{ g^2 \Tr T^a T^a}{(2 \pi)^{4}} \int_{0}^{1}dx\ \int d^4q\ \frac{(\slashed{\mathbf{p}}x-\slashed{\mathbf{q}})-3m_{3D}}{\left[\mathbf{q}^2+\mathbf{p}^2 x(1-x) +m^2_{3D}(1-x)+q_0^2x\right]^2}
\end{equation}
the denominator is now an even function in $\mathbf{q}$ and $q_0$ while we have an odd term on the nominator which is zero.\\
Now in order to have a convergent integral both in the UV and IR we use Pauli-Villars regularization on the gluon propagator
\begin{equation}
\frac{1}{q^2}\rightarrow \frac{1}{q^2+\mu^2}-\frac{1}{q^2+\Lambda^2}
\end{equation}
where we have put a small IR mass $\mu$ and a UV cutoff $\Lambda$. This amount to adding two terms to the formula for $\Sigma^{1\ loop}$ which becomes
\begin{align}
\Sigma^{1 loop}(\slashed{\mathbf{p}}) =&  -\frac{ g^2 \Tr T^a T^a}{(2 \pi)^{4}} \int_{0}^{1}dx\ \int d^4q\ (\slashed{\mathbf{p}}x-3m_{3D})\times\\\nonumber
&\Bigg[\frac{1}{\left[\mathbf{q}^2 +\mathbf{p}^2 x(1-x)+m^2_{3D}(1-x)+q_0^2x+\mu^2 x\right]^2}\\\nonumber
&-\frac{1}{\left[\mathbf{q}^2+\mathbf{p}^2 x(1-x) +m^2_{3D}(1-x)+q_0^2x+\Lambda^2 x\right]^2}\Bigg]
\end{align}
The integral in $d^4q$ can be performed in two steps, first $d^3\mathbf{q}$ and then $dq_0$. What we find when $\Lambda\rightarrow \infty$ is
\begin{align}
\Sigma^{1 loop}(\slashed{\mathbf{p}}) =  \frac{ g^2 \Tr T^a T^a}{16 \pi^2}\int_{0}^{1}dx\ \left({3m_{3D}-\slashed{\mathbf{p}}}x\right)\frac{1}{\sqrt{x}}\log\left(\frac{\Lambda^2 x}{x \left(\mu^2-\mathbf{p}^2 x+\mathbf{p}^2\right)-m^2_{3D} (x-1)}\right)
\end{align}
The integral is regular in $x$ due to the presence of the IR cutoff.  The divergent part of $\Sigma^{1 loop}(\slashed{\mathbf{p}})$ when the UV cutoff is sent to infinity turns out to be
\begin{align}
\Sigma^{1 loop}(\slashed{\mathbf{p}}) \sim  \frac{2 \alpha_s}{9 \pi} \left(9m_{3D}-\slashed{\mathbf{p}}\right)\log\left(\frac{\Lambda^2 }{m^2_{3D}}\right),\qquad \alpha_s = \frac{g^2}{4 \pi}
\end{align}
so the mass shift is log divergent
\begin{equation}
\delta m_{3D} \sim \frac{16 \alpha_s}{9 \pi} m_{3D}\log\left(\frac{\Lambda^2 }{m^2_{3D}}\right)
\end{equation}
while the field renormalization $Z^{-1}_{\varphi} =1- \frac{d \Sigma^{1 loop}(\slashed{\mathbf{p}})}{d \slashed{\mathbf{p}}}\big \rvert_{\slashed{p}=m_{3D}}$ is also logarithmically divergent. We have thus found that the parameters of the 3D action have the same divergent behaviour of the ordinary 4D fermions.
\bibliographystyle{JHEP}
\bibliography{mybibb} % BibTeX database without .bib extension

\providecommand{\href}[2]{#2}\begingroup\raggedright\begin{thebibliography}{10}

\bibitem{Parisi:1983ae}
G.~Parisi, \emph{{The Strategy for Computing the Hadronic Mass Spectrum}},
  \href{http://dx.doi.org/10.1016/0370-1573(84)90081-4}{\emph{Phys. Rept.} {\bf
  103} (1984) 203--211}.

\bibitem{Gusken:1989ad}
S.~Gusken, U.~Low, K.~H. Mutter, R.~Sommer, A.~Patel and K.~Schilling,
  \emph{{Nonsinglet Axial Vector Couplings of the Baryon Octet in Lattice
  {QCD}}}, \href{http://dx.doi.org/10.1016/S0370-2693(89)80034-6}{\emph{Phys.
  Lett.} {\bf B227} (1989) 266--269}.

\bibitem{Alexandrou:1990dq}
C.~Alexandrou, F.~Jegerlehner, S.~Gusken, K.~Schilling and R.~Sommer, \emph{{B
  meson properties from lattice QCD}},
  \href{http://dx.doi.org/10.1016/0370-2693(91)90219-G}{\emph{Phys. Lett.} {\bf
  B256} (1991) 60--67}.

\bibitem{Allton:1993wc}
{\scshape UKQCD} collaboration, C.~R. Allton et~al., \emph{{Gauge invariant
  smearing and matrix correlators using Wilson fermions at Beta = 6.2}},
  \href{http://dx.doi.org/10.1103/PhysRevD.47.5128}{\emph{Phys. Rev.} {\bf D47}
  (1993) 5128--5137}, [\href{https://arxiv.org/abs/hep-lat/9303009}{{\tt
  hep-lat/9303009}}].

\bibitem{Luscher:1990ck}
M.~{L\"uscher} and U.~Wolff, \emph{{How to Calculate the Elastic Scattering
  Matrix in Two-dimensional Quantum Field Theories by Numerical Simulation}},
  \href{http://dx.doi.org/10.1016/0550-3213(90)90540-T}{\emph{Nucl. Phys.} {\bf
  B339} (1990) 222--252}.

\bibitem{Basak:2005aq}
S.~Basak et~al., \emph{{Group-theoretical construction of extended baryon
  operators in lattice QCD}},
  \href{http://dx.doi.org/10.1103/PhysRevD.72.094506}{\emph{Phys. Rev.} {\bf
  D72} (2005) 094506}, [\href{https://arxiv.org/abs/hep-lat/0506029}{{\tt
  hep-lat/0506029}}].

\bibitem{Basak:2005ir}
{\scshape Lattice Hadron Physics (LHPC)} collaboration, S.~Basak et~al.,
  \emph{{Clebsch-Gordan construction of lattice interpolating fields for
  excited baryons}},
  \href{http://dx.doi.org/10.1103/PhysRevD.72.074501}{\emph{Phys. Rev.} {\bf
  D72} (2005) 074501}, [\href{https://arxiv.org/abs/hep-lat/0508018}{{\tt
  hep-lat/0508018}}].

\bibitem{Johnson:1982yq}
R.~C. Johnson, \emph{{Angular momentum on a lattice}},
  \href{http://dx.doi.org/10.1016/0370-2693(82)90134-4}{\emph{Phys. Lett.} {\bf
  B114} (1982) 147--151}.

\bibitem{Bruno:2014jqa}
M.~Bruno et~al., \emph{{Simulation of QCD with N$_{f} =$ 2 $+$ 1 flavors of
  non-perturbatively improved Wilson fermions}},
  \href{http://dx.doi.org/10.1007/JHEP02(2015)043}{\emph{JHEP} {\bf 02} (2015)
  043}, [\href{https://arxiv.org/abs/1411.3982}{{\tt 1411.3982}}].

\bibitem{Luscher:2011kk}
M.~Lüscher and S.~Schaefer, \emph{{Lattice QCD without topology barriers}},
  \href{http://dx.doi.org/10.1007/JHEP07(2011)036}{\emph{JHEP} {\bf 07} (2011)
  036}, [\href{https://arxiv.org/abs/1105.4749}{{\tt 1105.4749}}].

\bibitem{Luscher:2012av}
M.~Luscher and S.~Schaefer, \emph{{Lattice QCD with open boundary conditions
  and twisted-mass reweighting}},
  \href{http://dx.doi.org/10.1016/j.cpc.2012.10.003}{\emph{Comput. Phys.
  Commun.} {\bf 184} (2013) 519--528},
  [\href{https://arxiv.org/abs/1206.2809}{{\tt 1206.2809}}].

\bibitem{Bruno:2016plf}
M.~Bruno, T.~Korzec and S.~Schaefer, \emph{{Setting the scale for the CLS $2 +
  1$ flavor ensembles}},  \href{https://arxiv.org/abs/1608.08900}{{\tt
  1608.08900}}.

\bibitem{Blossier:2009kd}
B.~Blossier, M.~Della~Morte, G.~von Hippel, T.~Mendes and R.~Sommer, \emph{{On
  the generalized eigenvalue method for energies and matrix elements in lattice
  field theory}},
  \href{http://dx.doi.org/10.1088/1126-6708/2009/04/094}{\emph{JHEP} {\bf 04}
  (2009) 094}, [\href{https://arxiv.org/abs/0902.1265}{{\tt 0902.1265}}].

\end{thebibliography}\endgroup
\end{document}